\documentclass{aa}
\usepackage[varg]{txfonts}
\usepackage{mathptmx}
\usepackage{adjustbox,lipsum}

\usepackage{graphicx}
\usepackage{fixltx2e}
\bibpunct{(}{)}{;}{a}{}{,}

\def\Spitzer{{\it Spitzer}}
\def\GALEX{{\it GALEX}}
\def\uu{{\it u}}
\def\gg{{\it g}}
\def\rr{{\it r}}
\def\ii{{\it i}}
\def\zz{{\it z}}
\def\micron{~$\mu$m}

\newcommand{\angstrom}{~\textup{\AA}}

\begin{document}

\title{Spatially resolved star formation and dust attenuation in Mrk848:
  Comparison of the integral field spectra and the UV-to-IR SED}
\author{Fang-Ting Yuan\inst{1,2}, Mar\'{i}a
  Argudo-Fern\'{a}ndez\inst{3}, Shiyin Shen\inst{1}, Lei
  Hao\inst{1}, Chunyan Jiang\inst{1}, Jun
  Yin\inst{1}, M\'{e}d\'{e}ric Boquien\inst{3}, Lihwai
  Lin\inst{4}}

\institute{Key Laboratory for Research in Galaxies and
  Cosmology, Shanghai Astronomical Observatory, CAS, 80 Nandan Road,
  Shanghai 200030, China 
  \email{yuanft@shao.ac.cn}
  \and
  Aix-Marseille Universit\'e, CNRS, LAM (Laboratoire d’Astrophysique de
  Marseille) UMR7326, 13388, Marseille, France
  \and
   Universidad de Antofagasta, Unidad de Astronom\'ia, Facultad
   Cs. B\'asicas, Av. U. de Antofagasta 02800, Antofagasta, Chile
  \and
  Institute of Astronomy \& Astrophysics, Academia
  Sinica, Taipei 10617,Taiwan}

\date{Recieved,2017; accepted,??}

\abstract 
{We investigate the star formation history and the dust attenuation in
the galaxy merger Mrk848. Thanks to the multiwavelength photometry
from the ultraviolet (UV) to the infrared (IR), and MaNGA's integral
field spectroscopy, we are able to study this merger in a detailed
way. We divide the whole merger into the core and tail regions, and
fit both the optical spectrum and the multi-band spectral energy distribution (SED) to models to
obtain the star formation properties for each region respectively. We
find that the color excess of stars in the galaxy $E(B-V)_s^{\rm SED}$
measured
with the multi-band SED fitting is consistent with that estimated both
from the infrared excess (the ratio of IR to UV flux) and  from the
slope of the UV continuum. Furthermore, the reliability of the
$E(B-V)_s^{\rm SED}$ is examined with a set of mock SEDs, showing that
the dust attenuation of the stars can be well constrained by the
UV-to-IR broadband SED
fitting. The dust attenuation obtained from optical
continuum $E(B-V)_s^{\rm spec}$ is only about half of
$E(B-V)_s^{\rm SED}$. The ratio of the $E(B-V)_s^{\rm spec}$
to the $E(B-V)_g$ obtained from the Balmer decrement is consistent
with the local value (around 0.5). The difference between the results
from the UV-to-IR data and the optical data is consistent with the
picture that younger stellar populations are attenuated by an extra
dust component from the birth clouds compared to older stellar
populations which are only attenuated by the diffuse dust.  
Both with the UV-to-IR SED fitting and
the spectral fitting, we find that there is a starburst younger than
100~Myr in one of the two core regions, consistent with the
scenario that the interaction-induced gas inflow can enhance the star
formation in the center of galaxies.}  


\titlerunning{Spatially resolved star formation and dust attenuation in Mrk848}
\authorrunning{Yuan et al. (2017)}
\maketitle

\section{Introduction}

Galaxy mergers play an important role in the formation and
evolution of galaxies. Many studies have confirmed
  that the merging event is responsible for the elevated star formation
  activity in (ultra)luminous infrared  galaxies
  \citep{sanders1988,melnick1990}, and
the interaction between two galaxies can enhance the star formation
activity in these galaxies
\citep[e.g.,][]{barton2003,sanchez2003,alonso2004,ellison2008,xu2010,zhang,taniguchi,yuan2012,wild2014}.
However, the enhanced star formation rate (SFR) compared to
  non-interacting galaxies and its dependence on merging stages are
still quantitatively uncertain. Moreover, the spatial extent of star
formation in
interacting galaxies is still under debate. Some 
observational and theoretical studies have shown that the star formation 
triggered in mergers is widespread in tidal tails and bridges 
\citep[e.g.,][]{mirabel1998,alonso2000,elmegreen2006,renaud2014,wild2014},
whereas there are also works arguing that for most cases the
enhanced star
formation is restricted to the central region of the galaxy
 \citep{matteo2007,schmidt2013,moreno2015}. 
Besides the current SFR, the star formation history (SFH) of a merger
is also important for
constraining the merging stages and the effect of the
interaction. However, estimating the SFH of galaxies suffers from
several uncertainties. Especially, due to the complication of the
interaction, the SFH of mergers is difficult to reconstruct
\citep{pforr2012,conroy2013,buat2014,boquien2014,smith2015}. 

One of the uncertainties to determine the star formation properties is
the dust attenuation. In the optical wavelengths, the dust attenuation
is degenerated with the stellar age. It is better estimated when
introducing ultraviolet-to-infrared multiwavelength data, as the
ultraviolet (UV) and infrared (IR) bands
are more sensitive to the dust properties. Methods that measure the
dust attenuation include (1) fitting the stellar populations and the
dust attenuation together for the spectra or the spectral energy
distribution (SED), (2) estimating from the IR to UV
ratio (infrared excess, IRX) and (3) calculating from the slope
of the UV continuum
($\beta$). All these methods can give the attenuation for the
stars in galaxies. When there are IR data, the dust attenuation can be
estimated more reliably since the IR data provide additional constraints
on the dust emission. 

If the optical spectra are available, the Balmer
decrement can also be used to derive the dust attenuation. The Balmer
decrement gives the attenuation for nebular emission, and then the
dust attenuation for stars is taken as a factor
$f$ of that for the nebular emission. The classical value of $f$ for
galaxies is 0.44 \citep{calz2000}. However, recent studies have shown
that the value can vary from 0.44 to 0.93
\citep[e.g.,][]{kashino2013,kreckel2013,price2014,puglisi2016},
indicating that the
dust attenuation for stars estimated from Balmer decrement may be
underestimated. The complexity of dust attenuation in gas and stars
suggests that the commonly used one-component dust model in the stellar
population synthesis may not be suitable, and a more sophisticated
model of dust should be used. The two-component dust model proposed by
\citet{cf2000} contains a dust component for the diffuse interstellar
medium (ISM) where old stars are located and a second dust component
with larger optical depth for
birth clouds where young stars are born. The age-dependent dust
model has already been adopted in several works
\citep[e.g.,][]{brinchmann2004,noll2009,buat2011,wild2011,lofaro2017}. 
The typical time scale for a
birth cloud is about 10~Myr \citep{bs1980,cf2000}. For normal
galaxies, the stellar population younger than 10~Myr makes up only a
negligible fraction of the
total mass, and therefore the two-component dust model gives similar
results to the one-component dust model \citep{tojeiro2009}. However,
for more complicated
cases like starbursts/mergers, because the young stellar population is
not negligible,
and the young stellar population may take longer to escape from
the birth clouds \citep{silva1998,panuzzo2007}, using the
two-component dust model might be necessary.

With the development of observational  techniques, more and more data
with wavelength coverage from UV to IR
are available to constrain the star formation and dust attenuation
properties in galaxies at a spatially resolved level. Especially, the
development of the integral field unit (IFU) provides us with a large
amount of spatially resolved spectra in the optical band, which
contains much more information than either the photometry data or the
one dimensional spectra. Based on these data, some studies have been
done focusing on
the star formation properties and the
dust attenuation in interacting galaxies
\citep[e.g.,][]{rich2011,wild2014}. However,
a comparison between the results derived from the optical IFU
data and those from the photometric data (UV to IR) has yet to be
made. Also, previous studies only focus on the distribution of the
current SFR in interacting galaxies, but few attempts
have been
made to study the SFH for interacting galaxies.  

We report the dust attenuation and the SFH of the core
and tail parts of the merger Mrk848 using both the IFU spectra and the
UV-to-IR photometry data. Due to its merger nature, Mrk848 has been
extensively observed across multiwavelengths, including photometric
data from UV ({\GALEX}) to IR (\Spitzer).
It has also been observed by the SDSS-IV/MaNGA project
\citep[Mapping Nearby Galaxies at
APO,][]{bundy2015,drory2015,law2015,yan2016}. The existence of both
the photometric and IFU data of Mrk848 provides a unique
opportunity to derive spatially resolved dust attenuation
and SFH using different methods. Through the comparison of the results
from different methods, we attempt to decipher how the
calculation of the dust attenuation affects the estimation of the SFH,
and whether a two-component dust model is better for this case. The
derived SFH for the core and tail regions can also provide some
constraints on the spatial extent of the interaction-triggered star
formation and the time scale of the merging process. 

In Section~\ref{sec:data}, we present the
photometric and IFU data of Mrk848. In Section~\ref{sec:dustatt},
  we compare the dust attenuation derived from UV-to-IR photometric
  data and from the
optical IFU data, and discuss the
possible physical mechanism for the different results of these
methods. In Section~\ref{sec:sf}, we examine the SFH and SFR obtained
given different dust attenuation. We summarize the work in
Section~\ref{sec:summary}.

\section{Data}
\label{sec:data}

Mrk848 is a major merger of two galaxies with stellar masses $\log
(M_{*}/M_{\odot})=10.44$ and $10.30$ \citep{yang2007} in the nearby universe
($z=0.041$). Its morphologies, with obvious tidal tails, indicate
  strong interaction between the two galaxies.  In Figure 1, we divide Mrk848 into six regions according to its tail and
core features. The radius of each region is $5''$, which is
chosen according to the size of the cores. The center coordinates of
each region are listed in Table~\ref{tab:regioncoord}. We analyze the
star formation and dust properties using the IFU and broadband SED
data in each region. For all the data, the Galactic
extinction has been corrected using the SFD map
\citep{schlegel1998} and the Galactic extinction curve of
\citet{cardelli1989}.

\begin{figure}
\centering
\includegraphics[width=8cm]{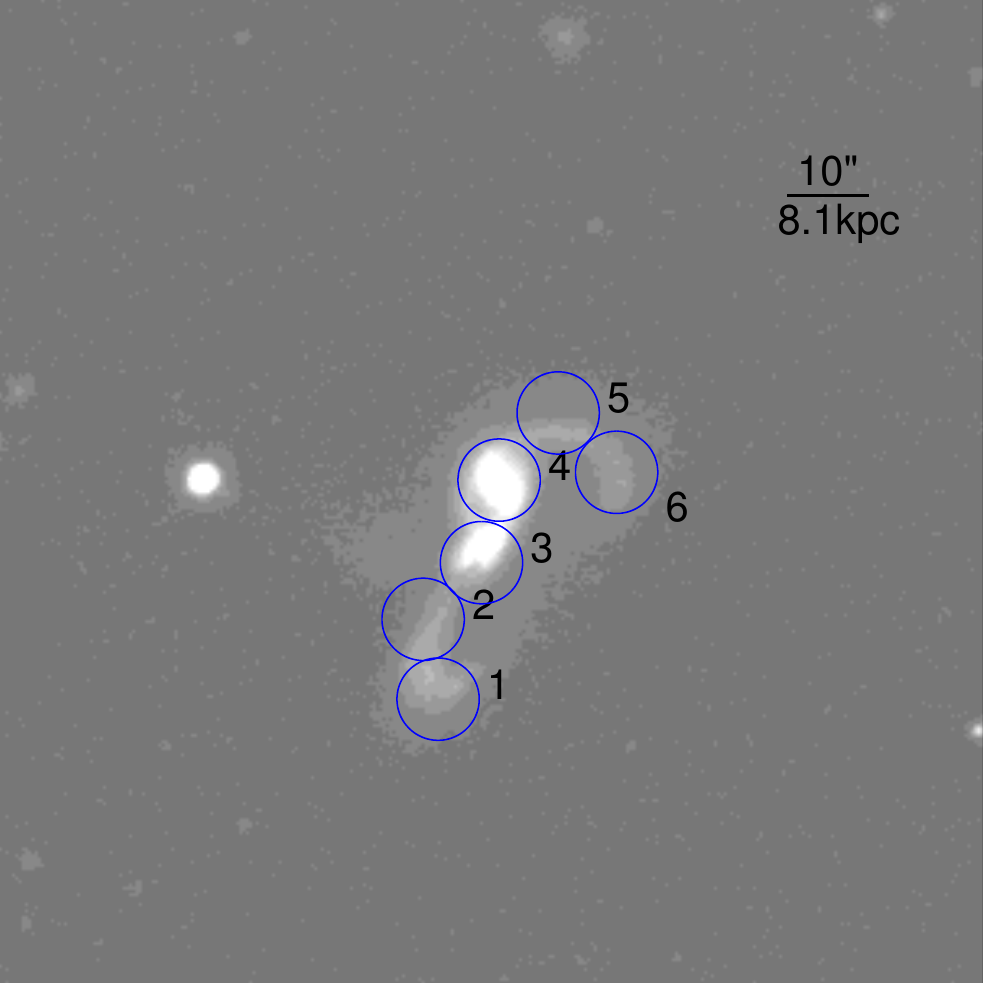}
\caption{Regions (cores: R3 and R4, tails: R1, R2, R5 and R6) of
  Mrk848 used for analysis in this work, overplotted on the \rr~band image
  of SDSS. The radius of each region is $5''$.  This image is rotated
  north-up.}
\label{fig:regions}
\end{figure}

\begin{table}
\centering
\caption{Center coordinates of each region and the signal-to-noise ratio (S/N) of the
  binned MaNGA spectra in each region.}
\begin{tabular}{cccc}
\hline\hline
Region  &  Center RA & Center Dec & S/N of MaNGA spectra\\
\hline
1  &  229.52836  &  42.738801 & \\
2  &  229.52906  &  42.741492 & 23.0\\
3  &  229.52636  &  42.743407 & 93.4\\
4  &  229.52556  &  42.746194 & 169.4\\
5  &  229.52285  &  42.748495 & 42.6\\
6  &  229.52015  &  42.746458 & 31.6\\
\hline
\end{tabular}
\label{tab:regioncoord}
\end{table}

\subsection{MaNGA integral field data}
\label{subsec:ifudata}

\begin{figure}

\centering
\begin{minipage}{3cm}
\includegraphics[width=0.8\linewidth]{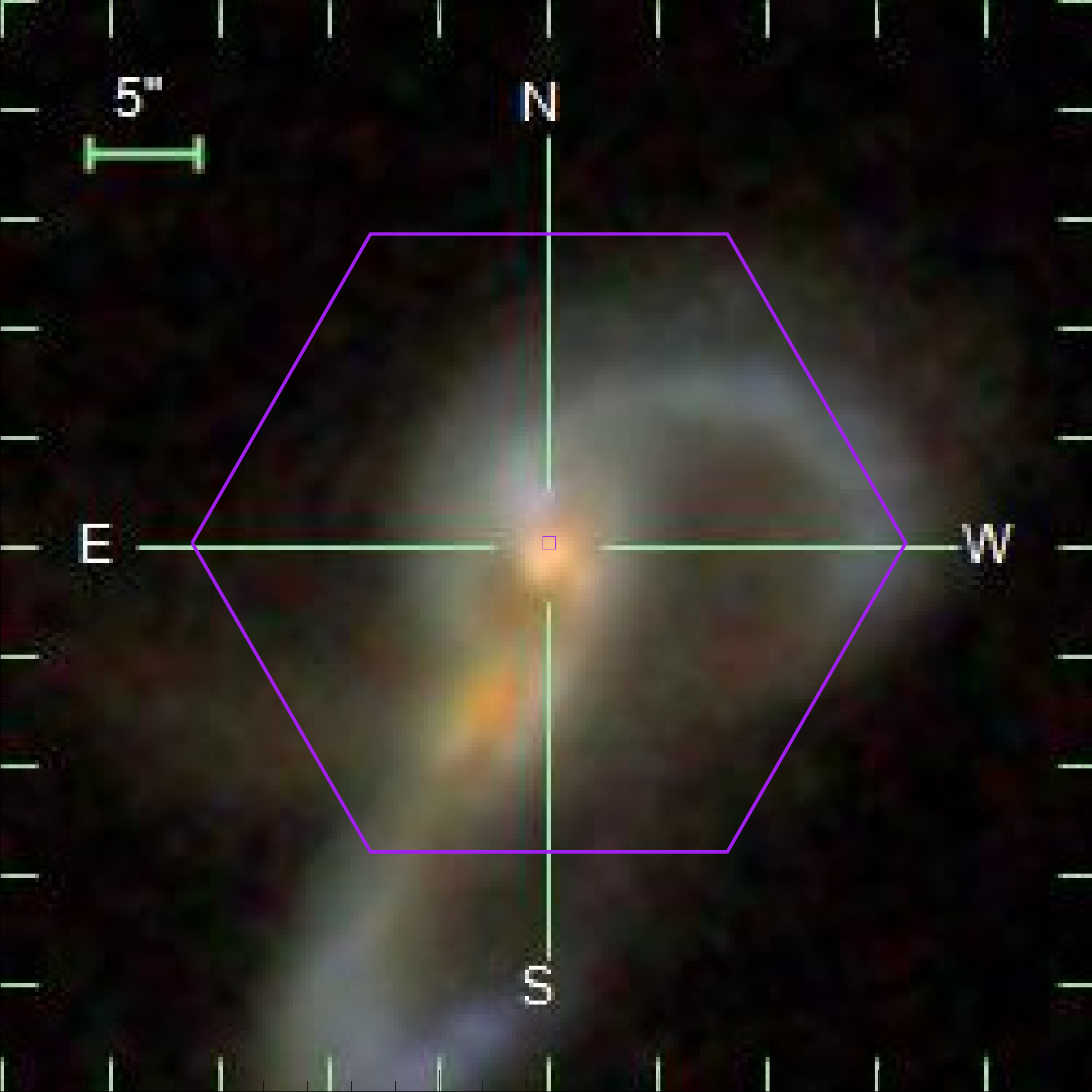}
\end{minipage}
\begin{minipage}{4cm}
\includegraphics[width=0.9\linewidth]{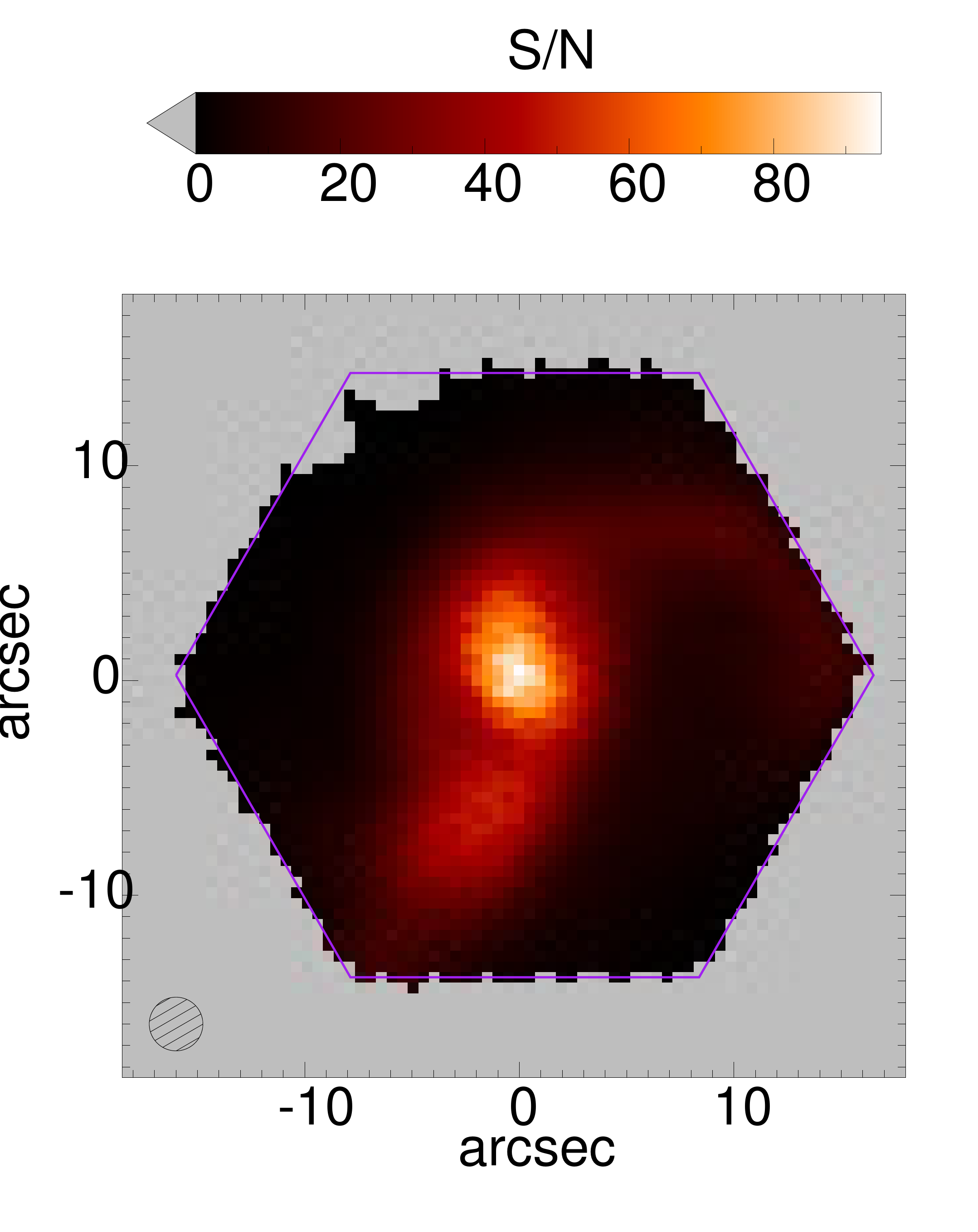}
\end{minipage}
\caption{Left panel: Coverage of MaNGA IFU bundle on Mrk848. Right
  panel: S/N distribution of MaNGA spectra.}
\label{fig:mangaobs}
\end{figure}

MaNGA is an IFU program to
survey 10,000 nearby galaxies using the BOSS spectrographs
\citep{smee2013} on the 2.5-meter SDSS telescope \citep{gunn2006,blanton2017}. The
wavelength coverage of MaNGA spectra is from 3600 to
10,000{\angstrom}, with a spectral resolution $R\sim2000$. The
diameters of MaNGA IFUs range from $12''$ to $32''$, composed of 19 to
127 fibers. The angular resolution (the FWHM of the fiber-convolved
PSF) of MaNGA data is about $2.5''$ \citep{yan2016}. Mrk848 (MaNGA
12-193481) is selected as one of the
commissioning targets of MaNGA
integral field spectra survey. It   
is observed by one of the largest IFU bundles with 127 fibers and a
diameter of $\sim32''$. With this size, the IFU fully covers R3,
R4, and R5, and
partially covers R2 and R6. R1 is not covered by the
IFU. The coverage of MaNGA IFU bundle is shown in Figure~\ref{fig:mangaobs}.

Figure~\ref{fig:mangaobs} also shows the S/N of each spaxel. 
For each region, the binned spectra are obtained by taking the 
average of all valid spaxels in that region. The errors are added
quadratically with a modifying 
term accounting for the covariance. The S/N after binning are
shown in Table~\ref{tab:regioncoord}.

\subsection{Multiwavelength photometric data}
\label{subsec:seddata}

\begin{figure}
  \includegraphics[trim={0cm 2cm 0cm 0cm},width=\linewidth]{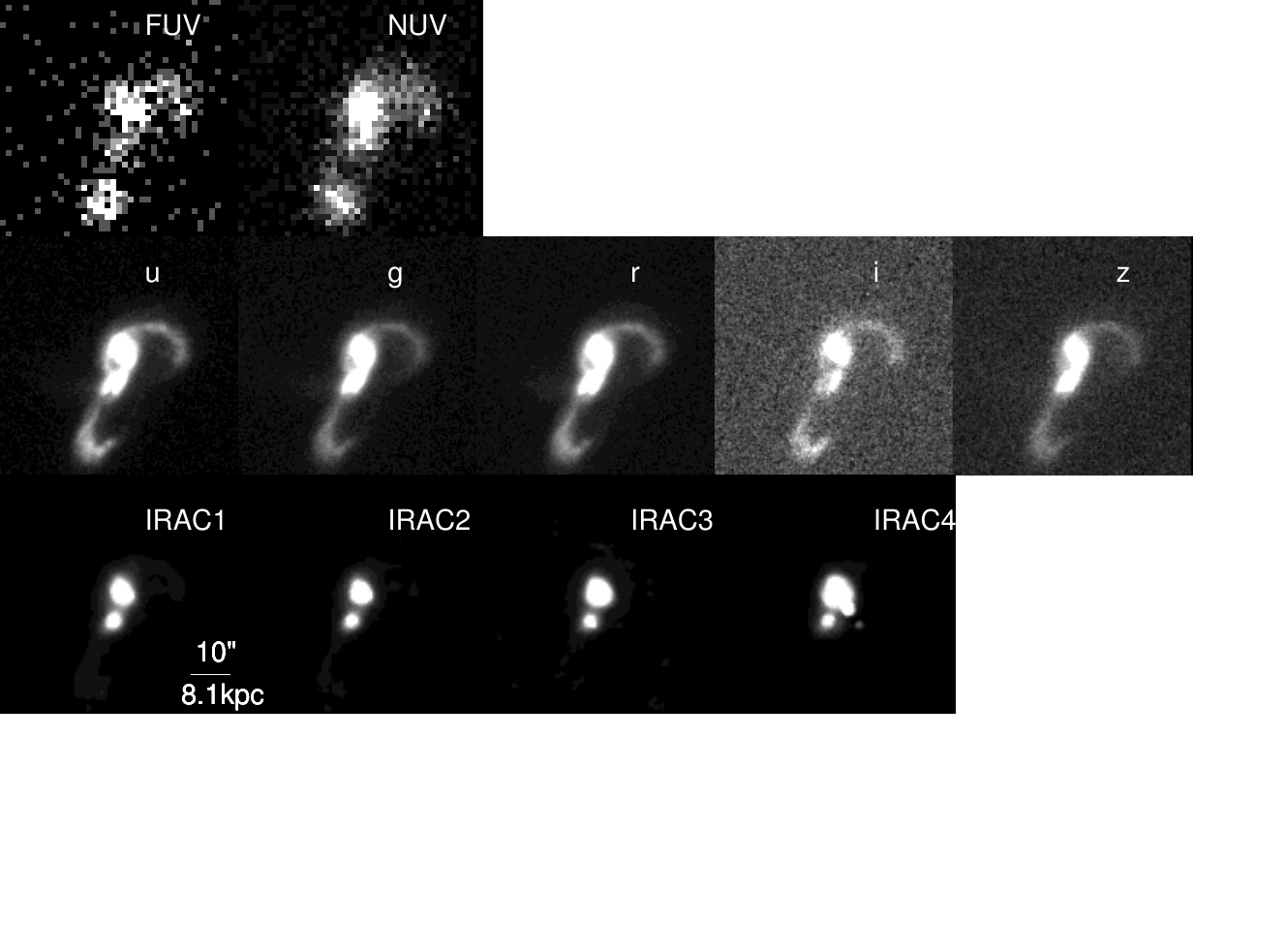}

  \caption{Images of Mrk848 from \GALEX, SDSS and \Spitzer,
    respectively. The size of each image is $1'\times1'$. All the images
    are rotated north-up.}
  \label{fig:images}
\end{figure}

We analyze the UV to IR photometric data to obtain the panchromatic
SED of each region. The observations, effective wavelengths, and
resolution of
the photometric data used in this work are
listed in Table~\ref{tab:observation}. The UV data are taken from
  {\GALEX} All
Imaging Survery \citep{martin2005}, including the FUV image at
{$1528$\angstrom} and the NUV image at $2271$\angstrom. The
resolution is typically 4.5/6.0 arcseconds (FWHM) for FUV/NUV band
\citep{madore2005}. 
At optical wavelengths, we take SDSS images at \uu~($\sim$
3551\angstrom), \gg~($\sim$ 4686\angstrom), \rr~($\sim$
6166\angstrom), \ii~($\sim$ 7480\angstrom), and \zz~($\sim$
8932\angstrom) bands. The spatial resolution of the SDSS images is
about $1.5''$. The values are estimated using the effective FWHM of
the PSF fitting by a double-Gaussian profile
\footnote{\url{http://www.sdss.org/dr12/imaging/images/#psf}}. For the
IR part, Mrk848 has been observed by Spitzer (AOR key is
12318720). The mean FWHM of the PSF for the {\Spitzer}/IRAC is about $2''$
\citep{fazio2004,aniano2011}. 
The longest wavelength used in
this work is {8\micron} from the IRAC observation. For even longer
wavelengths in IR, the two
interacting galaxies can not be spatially resolved. The images in
these bands are shown in Figure~\ref{fig:images}. 

\begin{table}

\caption{UV to IR images used in this work and their resolutions.}
\centering
\begin{tabular}{c c c}
\hline\hline
Observation   &    Wavelengths   &   Resolution \\
\hline
GALEX FUV     &    $1528$\angstrom &   4.5$''$  \\
GALEX NUV     &    $2271$\angstrom &   6.0$''$  \\
SDSS  \uu     &    3551\angstrom  & $\sim$ $1.5''$\\
SDSS  \gg     &    4686\angstrom  & $\sim$ $1.3''$\\
SDSS  \rr     &    6166\angstrom  & $\sim$ $1.3''$\\
SDSS  \ii     &    7480\angstrom  & $\sim$ $1.2''$\\
SDSS  \zz     &    8932\angstrom  & $\sim$ $1.2''$\\
{\Spitzer} IRAC1  &  3.6\micron  & $\sim$ $2''$\\
{\Spitzer} IRAC2  &  4.5\micron  & $\sim$ $2''$\\
{\Spitzer} IRAC3  &  5.8\micron  & $\sim$ $2''$\\
{\Spitzer} IRAC4  &  8\micron  & $\sim$ $2''$\\
\hline
\end{tabular}
\label{tab:observation}
\end{table}

\begin{figure*}
  \centering
  \includegraphics[width=\linewidth]{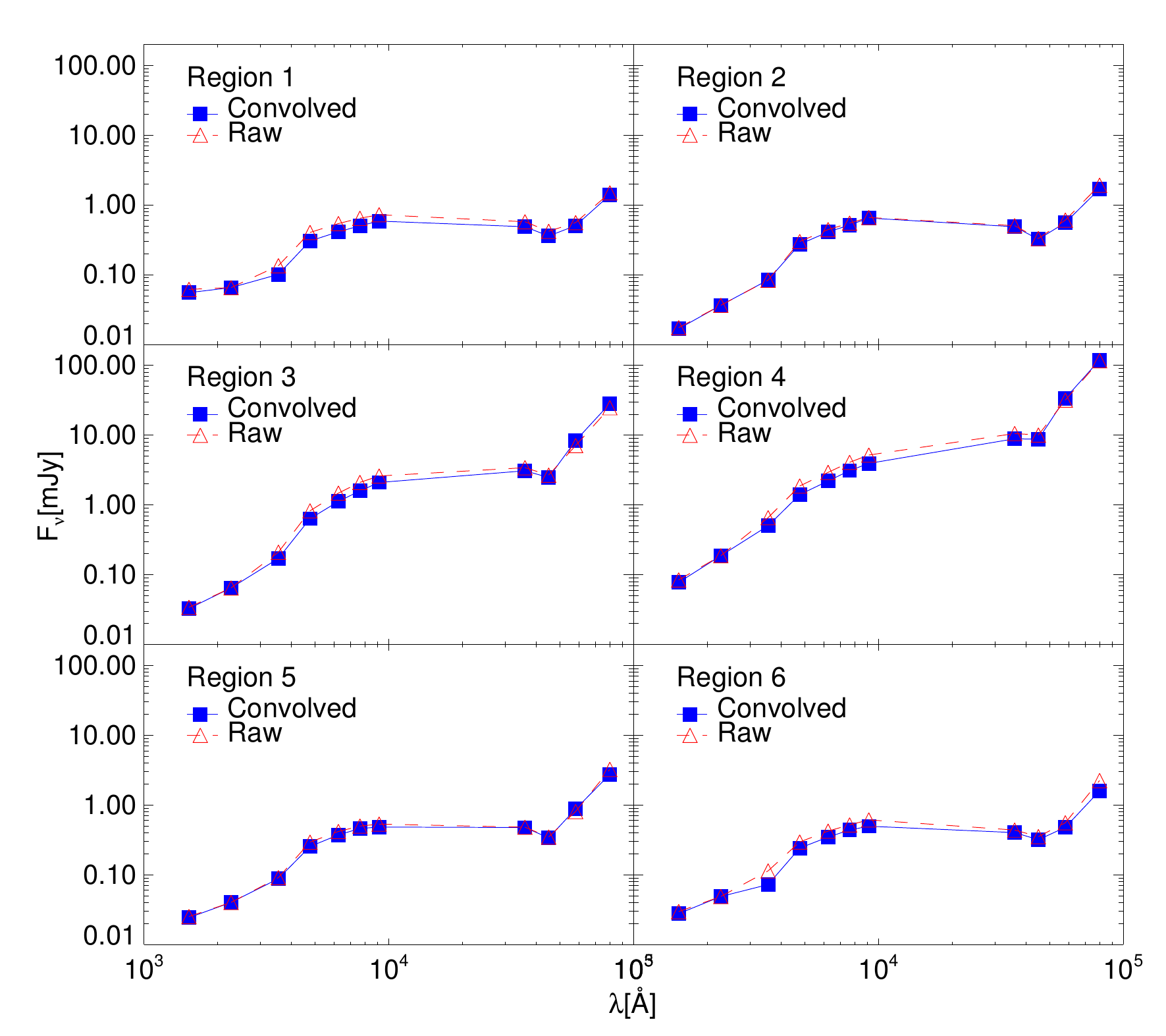}
  \caption{SEDs of the six regions in Mrk848. Triangles show
  SEDs derived from the images with their original resolution. Squares
  show SEDs derived from the images after degrading the resolution to
  $6''$ by convolving with the NUV PSF.}
  \label{fig:sed}
\end{figure*}

To obtain the UV-to-IR SED of each region, we calculate the fluxes in
each band within the radius of each region. We
estimate the background using an annulus $30''$ away from the center
with a width of $5''$. The flux of each region is obtained by summing up
the fluxes for the pixels in each region with the
background subtracted. The SEDs obtained from
the photometry are shown in Figure~\ref{fig:sed}.

Since the spatial resolution of these bands are inhomogeneous, to match the
photometry in different bands, the images of
each band should be converted into the same resolution. Here we use the
{\GALEX} NUV image (with the worst resolution, $6''$) as the reference.
   
The following steps are taken for these images:
\begin{enumerate}
\item Align all the images to WCS, and rescale the pixel size to the
  {\GALEX} NUV pixel size ($1.5''$) while keeping the surface
  brightness of the images unchanged.

\item Degrade all the images to the {\GALEX} NUV point spread function
  (PSF) using the large
  set of kernels presented by \citet{aniano2011}.
\end{enumerate} 

Then we calculate the SED of each region again. The results are
overplotted in Figure~\ref{fig:sed}. It appears that the difference
caused by the PSF in such a radius ($5''$) is small (the median
difference for each region is $<15$\%, Figure~\ref{fig:sed}). Hereafter,
we adopt the SEDs derived from the PSF-convolved images.

As shown in Figure~\ref{fig:sed}, the SED shapes of
the core regions (R3 \& R4) are apparently steeper than the tail
regions. The
infrared emission is much brighter in core regions than in tail
regions, indicating that the core regions have more active star
formation and heavier dust extinction (active galactic nuclei (AGN) contribution is
  negligible as we discuss in Section~\ref{subsec:agn}).

\section{Dust attenuation in Mrk848}
\label{sec:dustatt}

Since Mrk848 is very bright in the IR band (the total luminosity
in $8-1000${\micron} $L_{\rm TIR}>10^{12}L_{\odot}$), it is expected to have
high dust attenuation. To explore the stellar populations in each
region, we first examine their dust attenuation properties. We
  begin with the UV-to-IR broadband SED fitting. The long
  wavelength coverage of data can constrain the thermal emission by
  dust. In combination with the UV-optical data, these data provide
  strong constains on the dust attenuation. Then
  we compare the results with the dust attenuation derived from the
  $L_{\rm TIR}/L_{\rm FUV}$ ratio (infrared excess, IRX), the slope of
  the UV continuum ($\beta$), the optical continuum fitting and the
  Balmer decrement, respectively.


\subsection{UV to IR broadband SED fitting}
\label{subsec:sedfitting}

We use CIGALE (Code Investigating GALaxy Emission) to 
fit the multiwavelength SED from UV to IR. CIGALE can
fit the UV to IR data simultaneously by balancing the dust 
absorption in UV/optical bands and the dust emission in IR bands. 
This code has been widely used in literature to derive the star
formation and dust attenuation in galaxies from
the UV to IR multiwavelength SEDs \citep[e.g.,][]{giovannoli2011,buat2012}.  
A detailed description of the code can be found in \citet{noll2009}.

Here we describe the main features in our fitting. For the stellar
emission, we adopt the stellar population synthesis models of 
\citet{BC03} (hereafter BC03) and the Salpeter IMF \citep{salp}. We
consider a two-exponentially decreasing SFH, 
that is, an older stellar population with an exponentially decreasing SFH
added with a younger stellar population that indicates a later
starburst. The ages and $e$-folding time of the two components are free
parameters. The later burst can be used as an indication of the
merging-induced
star formation \citep[e.g.,][]{hernquist1989,torrey2012,patton2013}.
The two components are linked by the mass fraction of the burst
$f_{burst}$. The metallicity is fixed to solar metallicity during the
  SED fitting. Other metallicities have also been tested. The
  changes of the results caused by the metallicity are not significant
  and do not affect our conclusions.
\begin{table*}
  \centering
  \caption{Parameters used in the SED fitting.}
  \begin{tabular}{l c }
    \hline\hline
    Parameter & Value \\ 
    \hline
     &{Double exp. decreasing SFH}\\  
    \hline
    $age$ (Gyr) & 13    \\
    $\tau_{main}$  (Gyr) & 5.0, 7.0, 9.0, 11.0, 13.0\\
    $f_{burst}$         & 0.0001, 0.0005, 0.001, 0.005, 0.01, 0.05, 0.1, 0.5\\ 
    $age_{burst}$ (Myr)         & 10, 50, 100, 200, 500, 800, 1000\\ 
    $\tau_{burst}$  (Myr) & 500, 1000, 1500  \\
    \hline
    &{Dust attenuation}\\  
    \hline
    $E(B-V)_{young}$    &  0.1, 0.2, 0.3, 0.4, 0.5, 0.6,\\
    &   0.7, 0.8, 0.9, 1.0      \\
    $f_{att}$ & 0.5\\
    \hline
    &{Dust templates: \cite{dale2014}}\\  
    \hline
    $\alpha$    &       0.5, 1.0, 1.25, 1.5, 2.0, 2.25, 2.5  \\
    \hline
    \label{tab:cigale_param}
  \end{tabular}

\end{table*}

The dust attenuation is modeled assuming the Calzetti extinction
  curve with $E(B-V)_s$ as a free parameter. A reduction
factor of the visual attenuation, $f_{\rm att}$, is applied to the stars
older than 10~Myr to account for the distributions of stars of
different ages
\citep{cf2000,panuzzo2007,buat2012,lofaro2017}. Here we fix the
value of $f_{\rm att}$ to 0.5.
 
To simultaneously fit the IR part of the data, the dust emission
is modeled using the templates of \citet{dale2014}. The templates
are based on the star-forming-SED models modified with a certain fraction
of AGN contribution. Here we ignore the contribution from AGN
(See Section~\ref{subsec:agn}). The star-forming-SED models are
governed by the parameter $\alpha$, which relates the dust mass to
the heating intensity. The parameter $\alpha$ is an exponent
indicating the contribution of a series of
"local" SEDs representing the emission from dust
exposed to a wide range of heating intensities \citep{dh2002}. In
general, active star forming regions have a
smaller value of $\alpha$ than quiescent regions. The total
energy emitted by dust is estimated to balance the extincted energy
from the stellar emission.   

The main parameters and the input values for CIGALE are reported in 
Table~\ref{tab:cigale_param}. 
CIGALE creates models according to these input values and finds the
best-fit model by $\chi^2$ minimization. To probe a large
  parameter space while saving the computing time, we adopt a similar
  approach with \citet{buat2011}. We start with a
wide range of input parameters, and then reduce it by removing
values never chosen during the $\chi^2$ minimization. The output
values are estimated
by building the probability distribution function (PDF) and then taking 
the mean and standard deviation of the PDF. We derive the color
  excess of the stars $E(B-V)_{s}^{\rm SED}$ from $A_B-A_V$ calculated
  from the SED fitting. It is an
    effective $E(B-V)_{s}$ taking account of both the old stellar
    population and the young stellar population. The results of the
   $E(B-V)_{s}^{\rm SED}$ for each region in Mrk848 are listed in
  Table~\ref{tab:attres}.

\begin{table}
\centering
\caption{Color excess $E(B-V)_s$ in each region derived from different
  methods.}
\begin{adjustbox}{max width=\linewidth}
\begin{tabular}{c c c c c c c}
\hline\hline
Method & R1 & R2 & R3 & R4 & R5 & R6 \\
\hline
SED fitting & 0.114 & 0.207& 0.415& 0.484& 0.219&0.175\\
IRX & 0.100 & 0.193 & 0.374 & 0.497 & 0.195 & 0.153 \\
pPXF fitting & & 0.106 & 0.242 & 0.258 & 0.0834 & 0.0913 \\
\hline
\end{tabular}
\end{adjustbox}
\label{tab:attres}
\end{table} 

\begin{figure*} 
\centering
\includegraphics[width=0.95\linewidth]{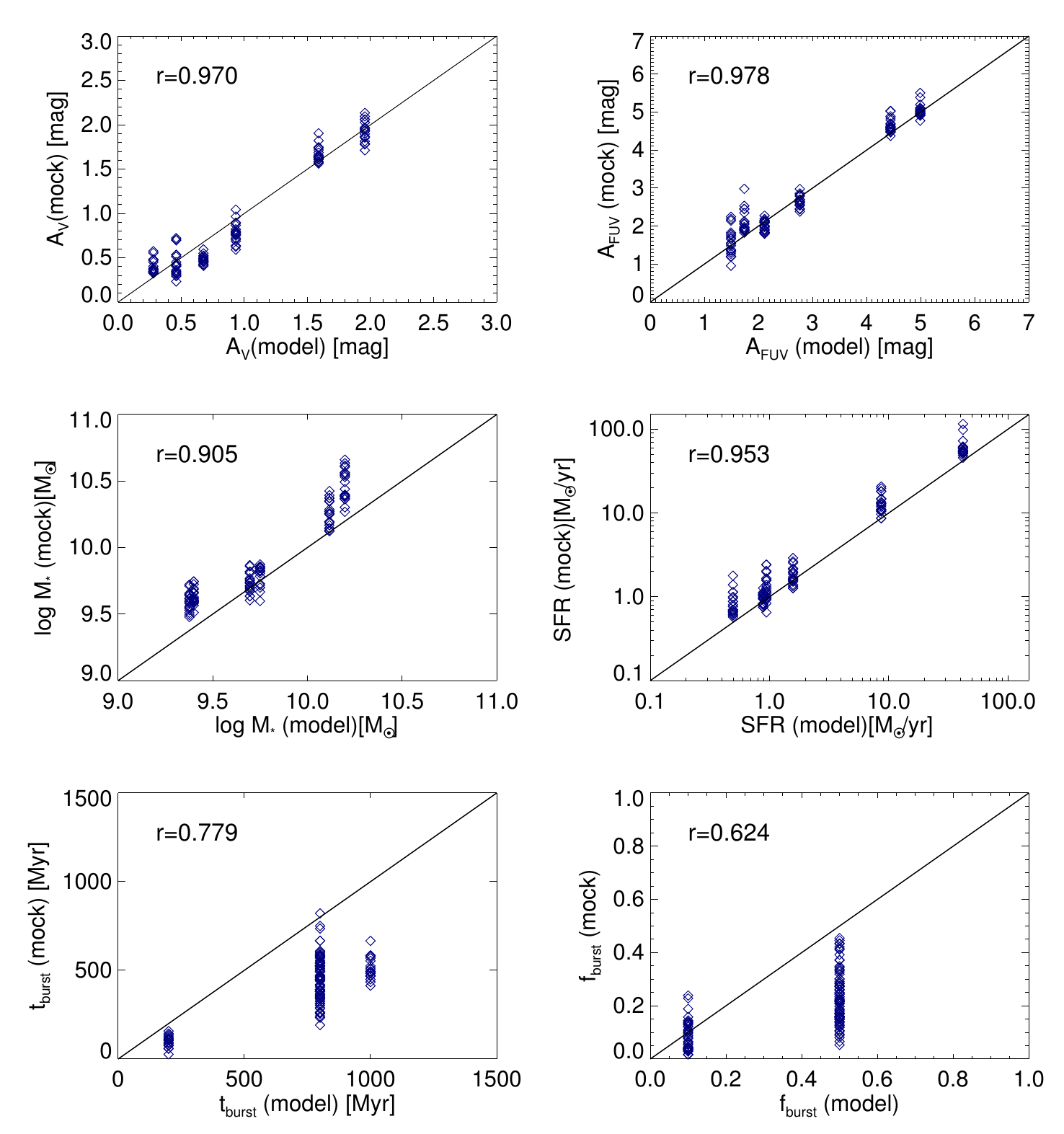}\\
\caption{Comparison of the values for six parameters of the best-fit
  models and their values estimated by CIGALE from the mock data based
  on the best-fit models. The mock results of the dust attenuation,
  the stellar mass and the SFR are scattered approximately around the
  best-fit parameters (correlation coefficient $r>0.9$), indicating
  that they are well constrained by the SED fitting. The burst
  fraction $f_{burst}$ and the burst age $t_{burst}$ are weakly constrained.} 
\label{fig:comp_mock}
\end{figure*} 
We use mock data to test how well the fitting can constrain the
  output parameters, with the method described by
\citet{giovannoli2011} and \citet{boquien2012}. The mock photometry in
each region is created from the best-fit
model of each region by deviating the best-fit model flux with a
random error $\sigma$, which follows Gaussian distribution with a standard
deviation of 10\% flux. For each region, we generate 20 sets of mock
SEDs. Then we do the SED fitting for the mock SEDs with the same input
parameters described above. The fitting
results are compared with the intrinsic values of the best-fit model
(Fig.~\ref{fig:comp_mock}). For the dust attenuation the mock
results are scattered approximately around our best-fit parameters
(correlation coefficient $r>0.9$), indicating that they are well
constrained by the SED fitting.

\subsection{Dust attenuation estimated from the UV/IR indicators}
\label{subsec:dustsed}
Although the fitting method is quite efficient at deriving several
physical parameters of galaxies at one time, it can bring
uncertainties caused by the degeneracy of parameters. It also
  depends on the assumptions of the SFH and the dust extinction of the
young and the old stellar populations. To investigate the
reliability of the fitting method, we calculate the dust
attenuation directly using two indicators. First, we
examine the $L_{\rm TIR}/L_{\rm FUV}$ ratio (infrared excess,
IRX), where $L_{\rm TIR}$ is the total infrared luminosity in
$8-1000${\micron}, and the $L_{\rm FUV}$ is the $\nu L_{\nu}$
calculated in the FUV band. 
The IRX is widely recognized to be a robust measurement of dust
attenuation. The IR and the UV emission together can provide good
constraints on the dust attenuation of galaxies. Here we adopt the
formula derived by \citet{buat2005}: 
\begin{equation}
A_{\rm FUV}^{\rm IRX}{\rm [mag]}=-0.0333x^3+0.3522x^2+1.1960x+0.4967,
\end{equation}
where 
\begin{equation}
x=\log\left(\frac{L_{\rm TIR}}{L_{\rm FUV}}\right).
\end{equation}
The IR luminosity $L_{\rm TIR}$ is estimated using the 8{\micron} flux.
We adopt the median value of the calibrations given by \citet{wu2005},
\citet{reddy2006}, \citet{bavouzet2008} and \citet{nordon2012}. The
standard deviation of these
calibrations is taken as the uncertainty of the value. Assuming that the
extinction obeys the Calzetti Law, the attenuation in the FUV band is
related to the color excess of the stars $E(B-V)_{s}^{\rm IRX}$ as 
\begin{equation}
A_{\rm FUV}=10.23E(B-V)_{s}^{\rm IRX}.
\end{equation}

\begin{figure}
\centering
\includegraphics[width=8cm]{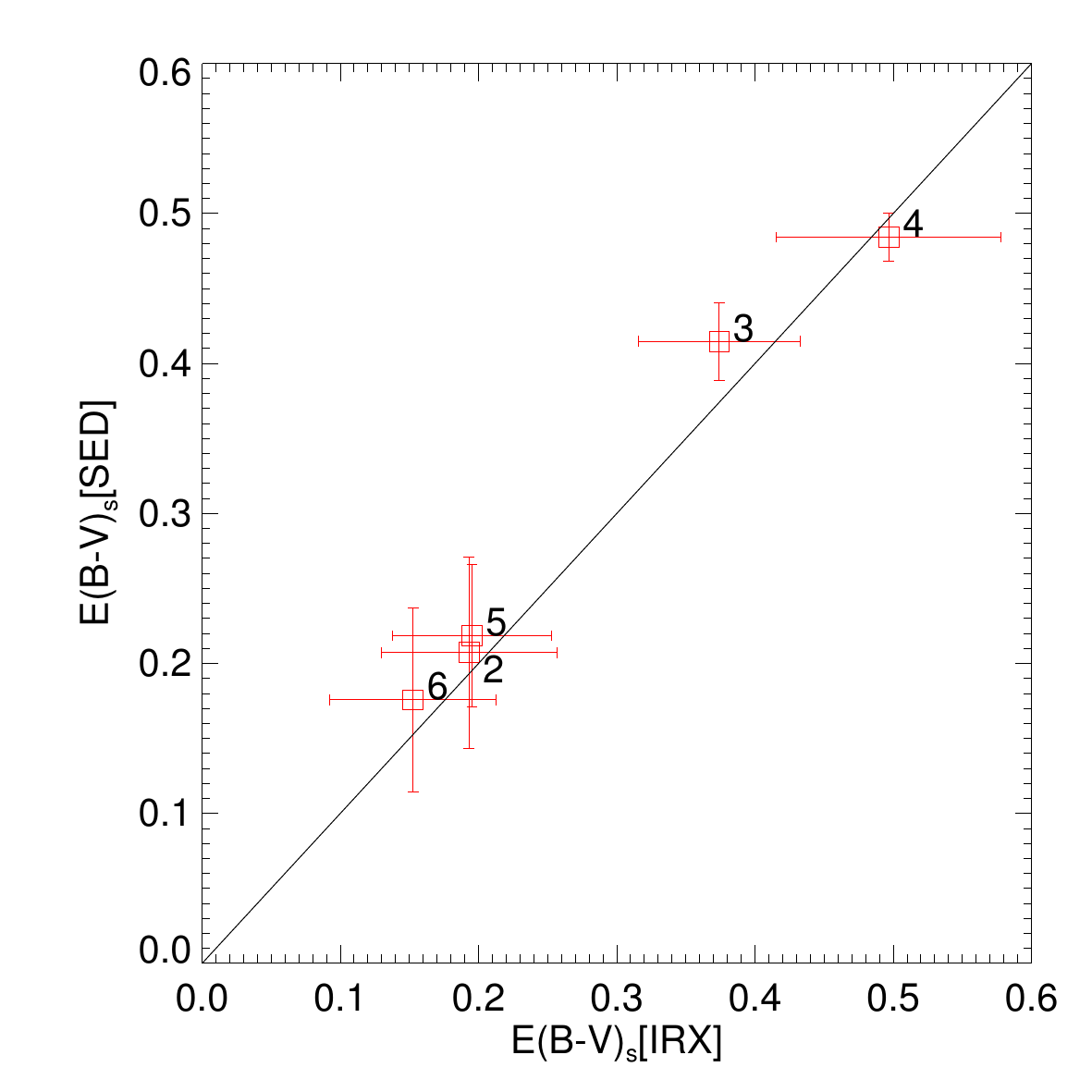}\\
\caption{Comparison of the $E(B-V)_s$ derived from the IRX with that
  derived from the SED fitting (squares). The line indicates the 1:1
  ratio.} 
\label{fig:compare_irx_fit}
\end{figure}

The results of $E(B-V)_s^{\rm IRX}$ are listed in
  Table~\ref{tab:attres}. The comparison between the
$E(B-V)_s$ derived from the IRX and those derived from the SED shows
that they are in agreement with each other
(Fig.~\ref{fig:compare_irx_fit}).
  
Another option to measure the dust attenuation is to use the slope of
the UV continuum. It is commonly used as a proxy to estimate dust
attenuation when the IR data are not available. Using the GALEX NUV and FUV
data, we can derive the slope $\beta$ from the formula given by
\citet{kong2004}: 
\begin{equation}
\beta=\frac{\log~f_{\lambda}^{\rm NUV}-\log~f_{\lambda}^{\rm FUV}}{\lambda^{\rm NUV}-\lambda^{\rm FUV}},
\end{equation}
where $f_{\lambda}^{\rm NUV}$ and $f_{\lambda}^{\rm FUV}$ are the NUV
and FUV fluxes, respectively (in ${\rm
  ergs~cm^2~s^{-1}~\angstrom^{-1}}$). 

\begin{figure}
\centering
\includegraphics[width=8cm]{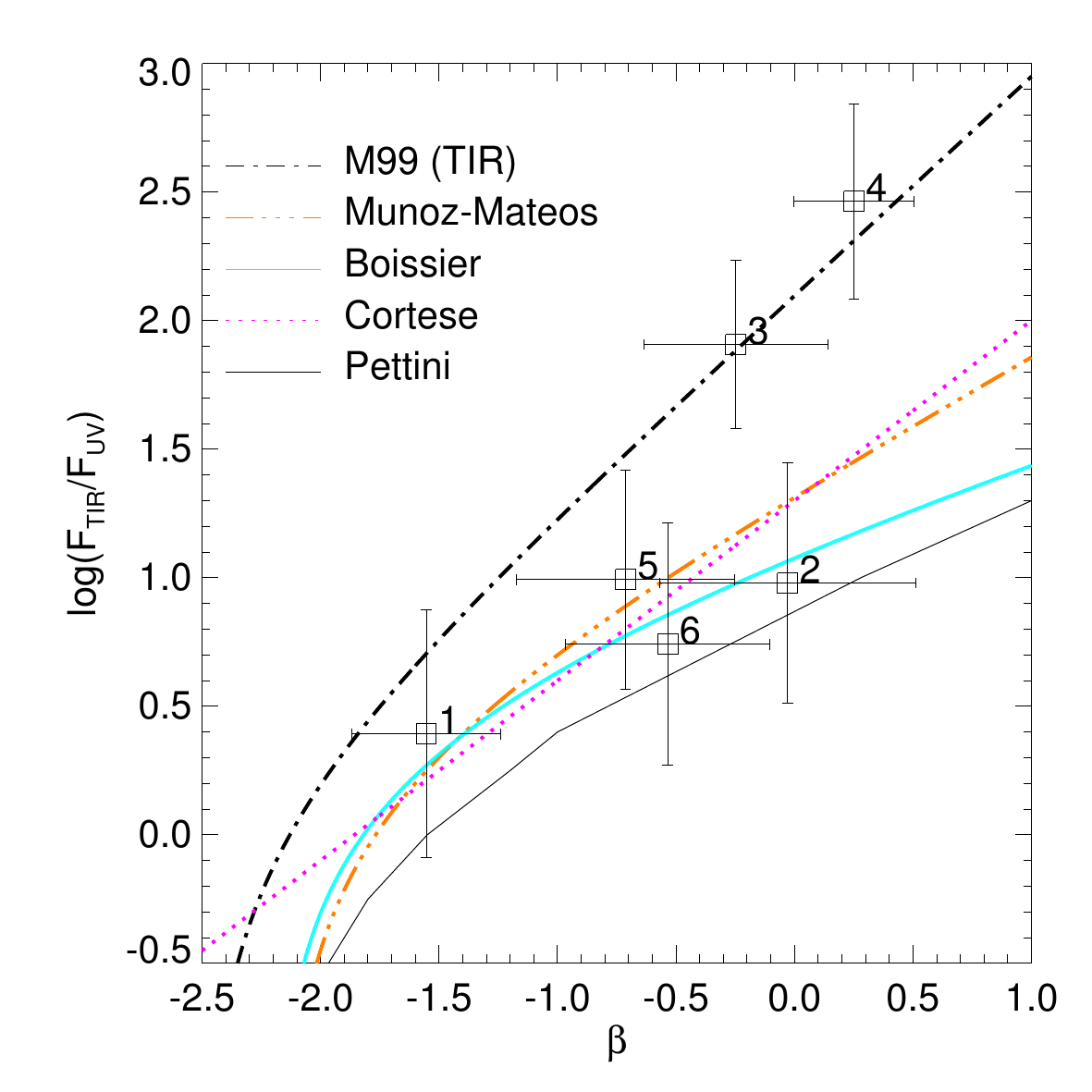}\\
\caption{IRX-$\beta$ relation for the six regions of Mrk848
  (squares with error bars). The
  relations given by \citet{meurer1999}, \citet{munoz-mateos2009},
  \citet{boissier2007}, \citet{cortese2006} and \citet{pettini1998}
  are overplotted. } 
\label{fig:comp_irx_beta}
\end{figure}

Assuming the IRX-$\beta$ relation given by \citet{meurer1999}, $\beta$ is
related to $A_{\rm FUV}$ as  
\begin{equation}
A_{\rm FUV}^{\beta}=4.43+1.99\beta.
\label{equ:m99}
\end{equation}
Recent studies have shown that this relation has a large
dispersion for different types of galaxies
\citep[e.g.,][]{kong2004,buat2005,burgarella2005,cortese2006,boissier2007,boquien2012,takeuchi2012}. 
A blind use of this relation will cause either underestimation or
overestimation of
the dust attenuation. Figure~\ref{fig:comp_irx_beta} shows the
IRX-$\beta$ relation for the six regions of Mrk848. The core regions
(R3 and R4) obey the relation derived from starbursts by
\citet{meurer1999}, while the IRX-$\beta$ relation of the tail regions
are more similar to that of galaxies with more quiescent star formation
activities. If using Equation~\ref{equ:m99}, the dust attenuation
derived from the UV slope will be larger than that derived from the
IRX. However, considering the variation of the IRX-$\beta$ relation
and using the IRX-$\beta$ relation given by
\citet{munoz-mateos2009}, \citet{boissier2007}, \citet{cortese2006}
or \citet{pettini1998} for the tail regions,
the attenuation derived from $\beta$ is consistent with that derived
from the IRX. 

\subsection{Dust attenuation estimated from fitting the optical continuum}
\label{subsec:specfitting}

The dust attenuation of each region can also be obtained by fitting the
optical spectra using stellar population synthesis models. We use pPXF
\citep[penalized Pixel-Fitting,][]{ce2004,cappellari2017} 
where the maximum penalized likelihood approach is adopted. We
choose 96 single stellar population (SSP) models of BC03 assuming a
Salpeter IMF \citep{salp}, with metallicity
[Z/H] ranging from -2.3 to 0.4, and ages ranging from 10~Myr to
13~Gyr. To obtain the dust attenuation, we assume the
Calzetti extinction curve \citep{calz2000} and allow pPXF to fit the
stellar color excess $E(B-V)_s$ as a free parameter together with
the kinematics and the weights of each SSP component.
The emission lines are masked during the fitting.

\begin{figure}
\centering
\includegraphics[width=8cm]{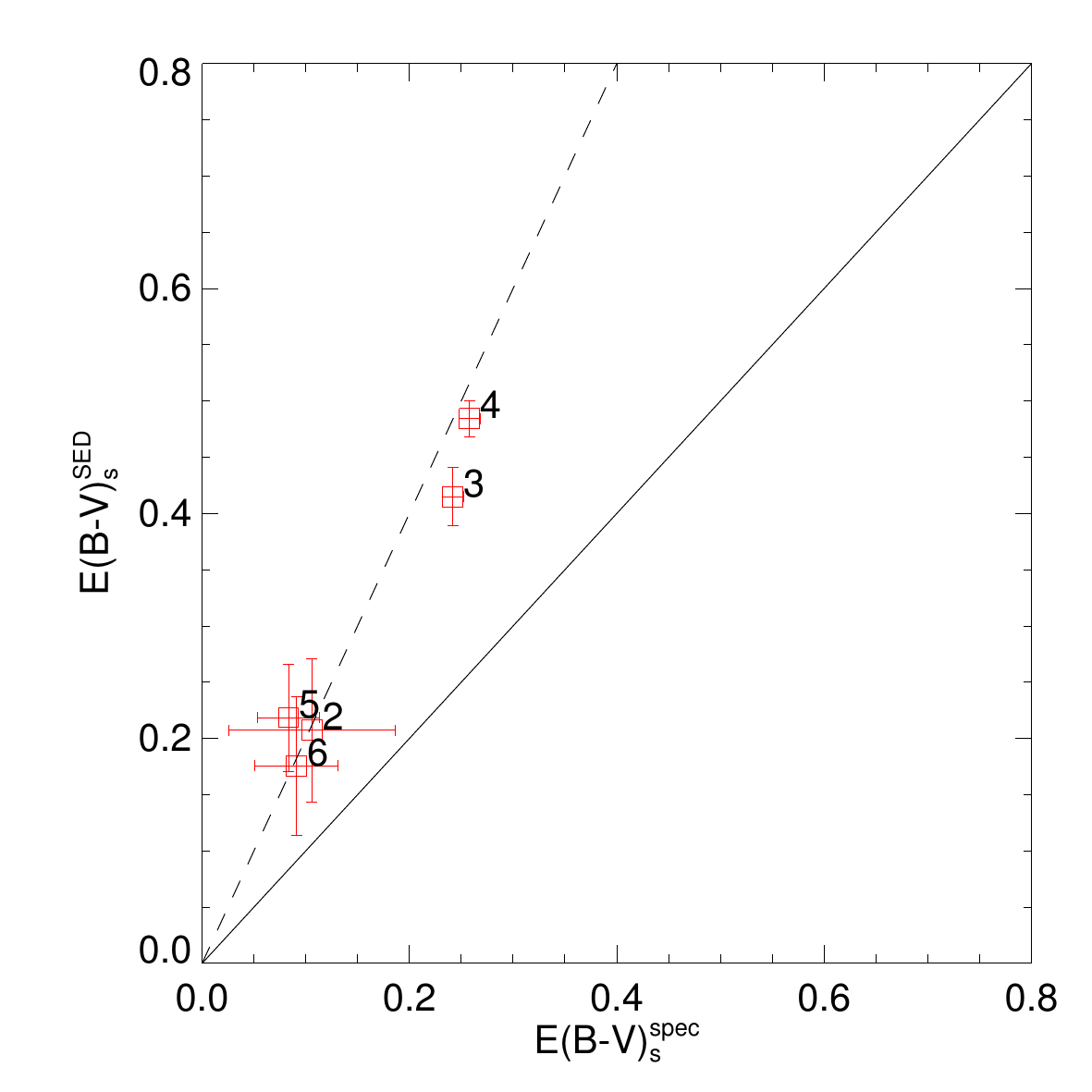}\\
\caption{Comparison of the $E(B-V)_s$ derived from the CIGALE SED
  fitting and the pPXF optical spectral fitting. The solid line is the
  1:1 line. The dashed line shows the best-fit that
  $E(B-V)_s^{\rm spec}$=0.5$E(B-V)_s^{\rm SED}$.}
\label{fig:compare_ebv}
\end{figure}

The results of the color excess $E(B-V)_s^{\rm spec}$
are listed in Table~\ref{tab:attres}. We find that
the $E(B-V)_s^{\rm spec}$ given by the pPXF optical spectral fitting
is a factor of 0.5 of the $E(B-V)_s^{\rm SED}$ 
given by the UV-to-IR SED fitting (Fig.~\ref{fig:compare_ebv}). This
difference indicates that the optical continuum cannot probe the
heavily attenuated regions as deep as the UV-to-IR data (See
Sect.~\ref{subsec:physinterp} for further discussion).

\subsection{Dust attenuation derived from the Balmer decrement}

With MaNGA spectra, we can obtain the dust attenuation from the Balmer
decrement, ${\rm H}\alpha/{\rm H}\beta$. The fluxes of emission lines
can be derived from MaNGA spectra after removing the continuum by
stellar population synthesis. Comparing the observed ratio 
${{\rm H}\alpha}/{{\rm H}\beta}$ with the theoretical value of 2.86
obtained for the Case B recombination, we can derive the extinction
$A({\rm H}\alpha)$ and the nebular color excess $E(B-V)_{g}$: 
\begin{equation}
E(B-V)_g=\frac{2.5}{k({\rm H}\beta)-k({\rm
    H}\alpha)}\log\left[{\frac{({\rm H}\alpha/{\rm
        H}\beta)_{obs}}{2.86}}\right]. 
\end{equation}
Assuming the reddening curve $k(\lambda)$ of \citet{calz2000}, the
value of $k({\rm H}\beta)-k({\rm H}\alpha)$ is 1.27. 

The color excess
of the nebular emission lines, $E(B-V)_{g}$, is linked to that of the
stellar continuum, $E(B-V)_{s}$, by a factor $f$, $E(B-V)_{s}=f
E(B-V)_{g}$. The classical value of $f$ derived from local galaxies is
$0.44$ \citep{calzetti1997,wuyts2011}. The value was first
  derived by \citet{calzetti1997} by comparing the UV slope to the
  $E(B-V)_g$ derived from H$\alpha$ for a sample of starburst
  galaxies, and then widely used in literature.
  \citet{wuyts2011} investigated a larger sample with data from
  UV-to-IR, and confirmed that this value is valid when
  correcting the SFR$_{{\rm H}\alpha}$ to
  SFR$_{\rm~UV+IR}$. \citet{pannella2015} show that $f$ should
be 0.58 if the $E(B-V)_s$ is derived from the Calzetti extinction
curve rather than the extinction curve of
\citet{fitzpatrick1999}. Works on high-z galaxies give larger $f$
values \citep[e.g.,][]{kashino2013,koyama2015, puglisi2016}. 

\begin{figure}
\centering
\includegraphics[width=8cm]{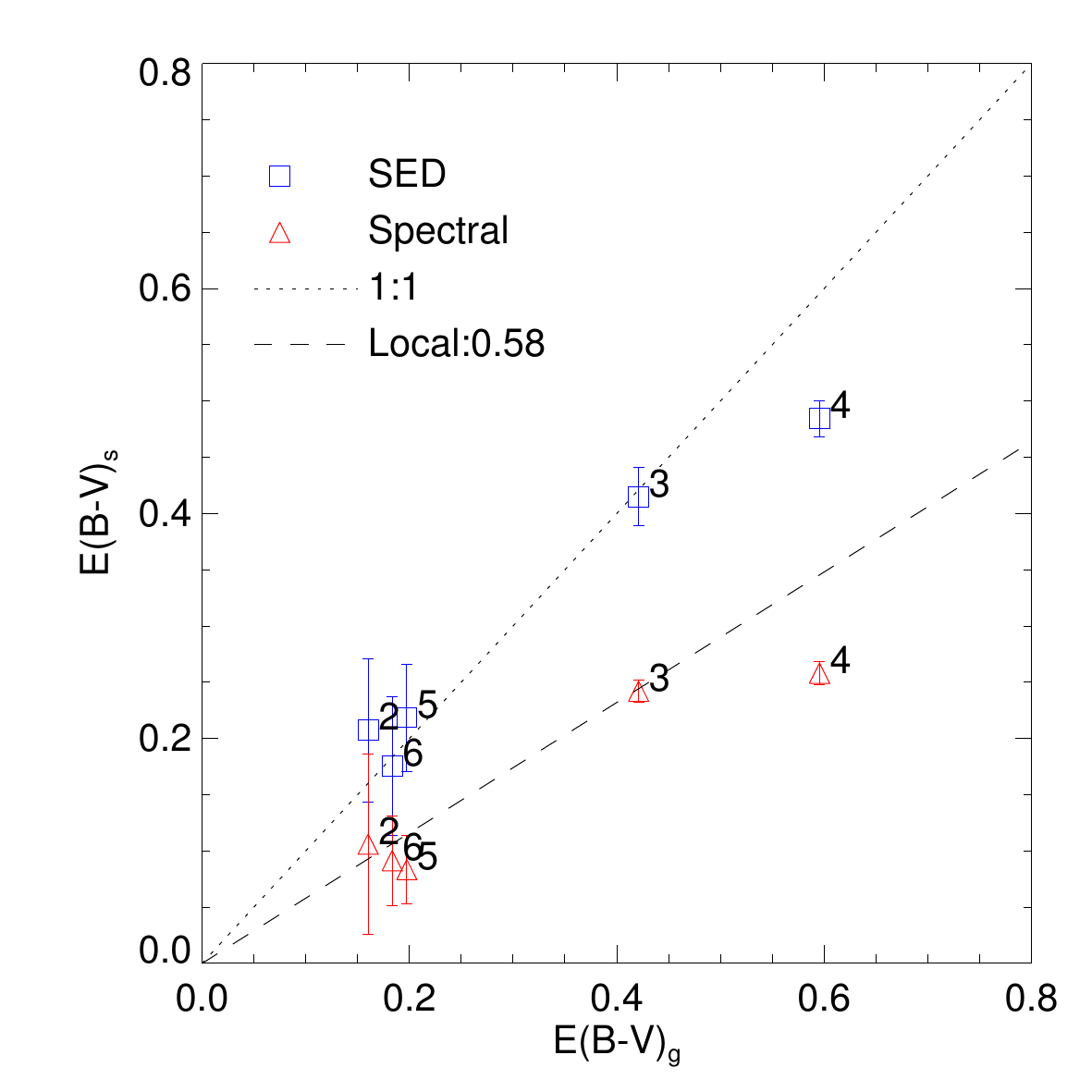}\\
\caption{Comparison of the $E(B-V)_g$ derived from the Balmer
  decrement with the $E(B-V)_s$ derived from the SED fitting (squares)
  and the spectral fitting (triangles). The dotted line indicates the 1:1
  ratio. The dashed line shows the local value (0.58) of
  $E(B-V)_s/E(B-V)_g$ given by \citet{pannella2015}.}
\label{fig:compare_ebv_gasstar}
\end{figure}

In Fig.~\ref{fig:compare_ebv_gasstar} we plot the $E(B-V)_s$ of the
regions of Mrk848 derived from the
SED fitting and the spectra fitting compared with the $E(B-V)_g$. The
SED fitting gives $f\sim1$ while the spectral
fitting gives $f\sim0.5$. The physical mechanism that may explain the
varying f factor is discussed in the following Section.

\subsection{Physical interpretation}
\label{subsec:physinterp}

From the above analysis, we find that the stellar color excess derived
from the SED fitting $E(B-V)_s^{\rm SED}$ is larger than that derived from the
optical spectra fitting $E(B-V)_s^{\rm spec}$, especially in the core regions
(R3 and R4). We used the mock data
  (Sect.~\ref{subsec:sedfitting}) and tested that the $E(B-V)_s^{\rm
    SED}$ are well constrained. We also showed in
Section~\ref{subsec:dustsed} that the $E(B-V)_s^{\rm SED}$ is
consistent with that obtained using the IR/UV indicators.
  However, using
  the $E(B-V)_g$ given by the Balmer decrement, the results from SEDs
  put $E(B-V)_s^{\rm SED}/E(B-V)_g$ to have a ratio of $\sim1$, which
  is inconsistent with the classical $f$ value of 0.44 
  \citep[or the local value 0.58 given by][]{pannella2015}. 
  On the other hand, the $E(B-V)_s^{\rm spec}$ to
   $E(B-V)_g$ ratio is 0.5, consistent with the the classical $f$
  value.

The difference among these results can be explained by the
two-component dust model
\citep[e.g.,][]{calz1994,cf2000,wild2011,price2014}, corresponding
to the similar implementation in our broadband SED fitting. As
illustrated in \citet{cf2000}, this model contains a diffuse dust
component in the interstellar medium (ISM) and a dust component in the
stellar birth clouds. The former affects both the old diffuse stellar
populations and the young stellar populations, whereas the latter
affects only the young stellar populations which are still embedded in
the birth clouds. Since Mrk848 is a merger, the gas inflow into the
center would lead to starbursts, making it necessary to consider the
two-component dust model. For galaxies with high specific SFRs
(SFR/$M_*$) as in our case (see Table~\ref{tab:sedres}), both the
emission lines and the UV-to-IR continuum emission, which is
dominated by the recently formed stars in the birth cloud, are
attenuated by both dust components. Therefore, the attenuation of
emission lines and that of the stars dominating UV-to-IR SED
would be similar.

In the framework of the two-component dust model, the
$E(B-V)_s^{\rm spec}$ would be lower than $E(B-V)_s^{\rm SED}$. In the
stellar birth cloud, the optical radiation
from both stars and gas will mostly be absorbed by the dense cloud
before escaping from the galaxy. Therefore, what we observed in
  optical bands will be dominated by the diffuse radiation from older
  stars in the diffused regions outside of the birth clouds. In
  contrast, the UV and IR SED detects emissions from more attenuated
  parts inside the birth clouds. Therefore, the two quantities
measure different dust attenuation.

\citet{wild2011} and \citet{price2014} suggest that the star to
  gas attenuation ratio is larger in higher SSFR galaxies, in
  agreement with our interpretation above - that the UV-to-IR SED
  detects dusty, high SSFR parts of Mrk848, whereas the optical
  continuum is dominated by the emission in the diffuse, low SSFR
  region. 

\begin{figure}
\centering
\includegraphics[width=8cm]{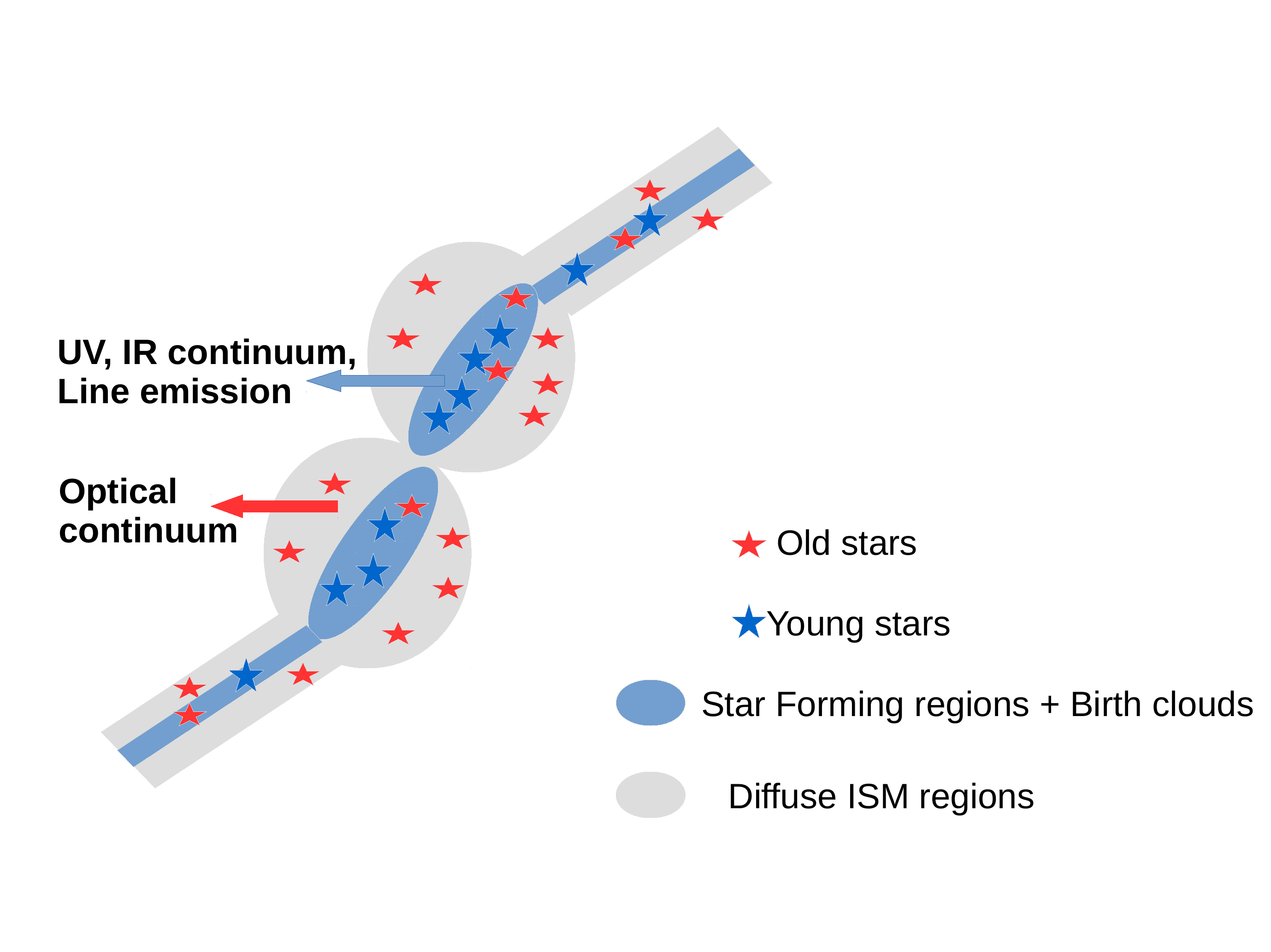}\\
\caption{Pictorial representation of the distribution of dust and
  stars in Mrk848. The optical continuum observed
is dominated by the diffuse region outside of the birth
cloud. The UV/IR continuum and the line emissions are dominated by the star
formation in the birth cloud that are more opaque than the diffuse
region. Therefore, the UV/IR continuum and the line emissions are more
attenuated than the optical continuum.}
\label{fig:toymodel}
\end{figure}

It is interesting that the tail regions also show some discrepancy
between optical- and UV/IR-derived dust attenuation, implying similar
dust geometry in the tail regions. A possible explanation is that
there are channels in the tail regions from which gas inflows to the
center, and during the inflow path, the gas also cools down and forms
stars. Figure~\ref{fig:toymodel} illustrates the physical picture
described above. We note that the actual situation may be more
complicated than this simple picture (see Sect.~\ref{subsec:extcurve}).

\begin{figure*}
\centering
\includegraphics[width=0.8\linewidth]{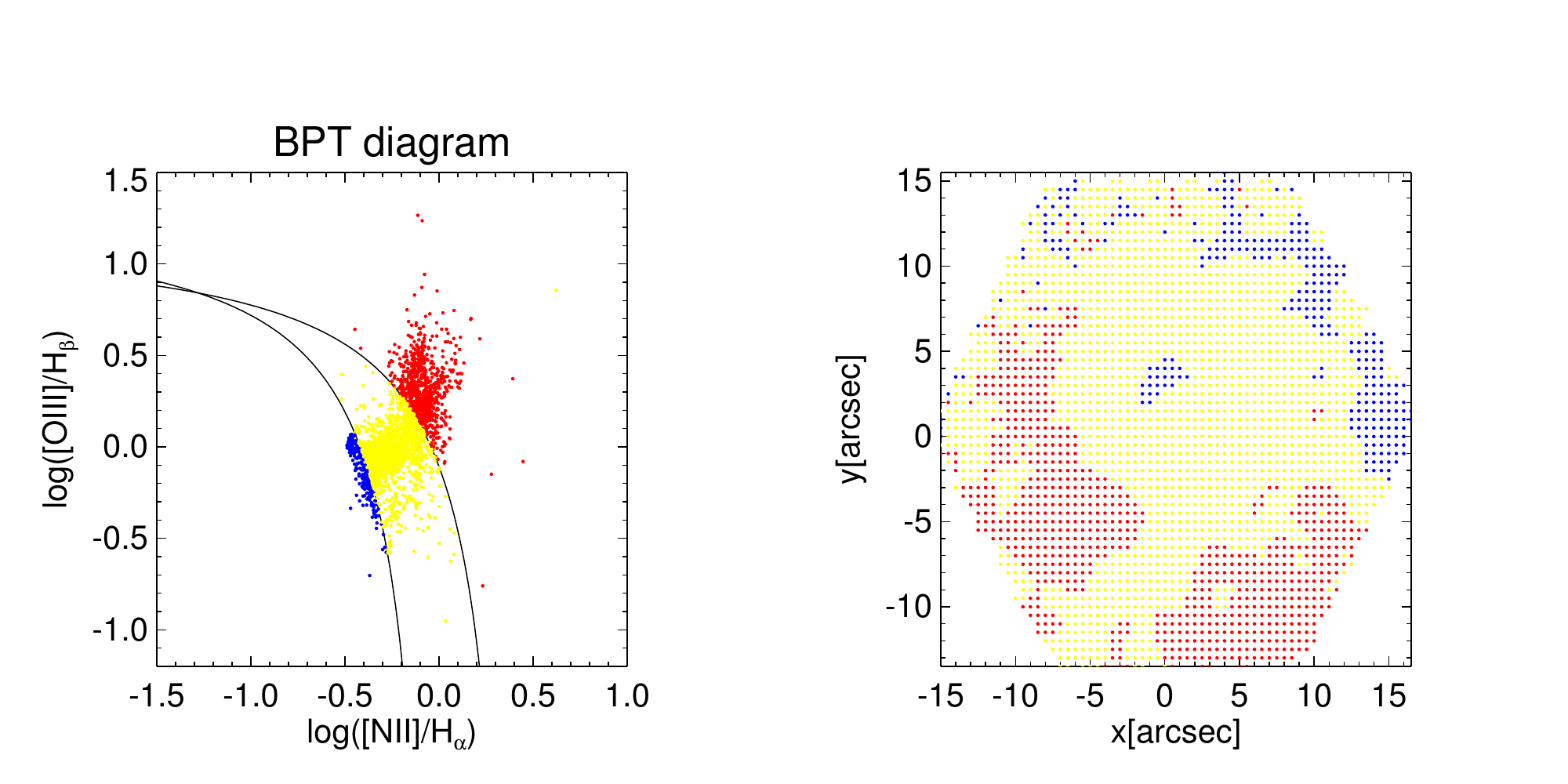}
\caption{Left: BPT diagram of Mrk848. Right: spatial
  distribution of the AGN (red), composite (yellow), and star-forming
  (blue) regions.}
\label{fig:bpt}
\end{figure*}

\subsection{Active galactic nuclei contribution}
\label{subsec:agn}

The X-ray observation of Mrk848 shows that it may encompass an AGN
\citep{brightman2011}.  We examine the emission lines obtained from
MaNGA spectra, and find that the equivalent widths of the permitted
emission lines (e.g., H$\alpha$ and H$\beta$) are less than
$300\,\rm{km\,s^{-1}}$. Therefore, the embedded AGN is Seyfert-2
type. The BPT diagram \citep{bpt} shows this system is dominated by
composite regions (Fig.~\ref{fig:bpt}), implying that the AGN
contribution to emission lines is not significant. The contribution to the 
continuum should be even less.
We also examine the AGN contribution using the SED fitting
with CIGALE. We adopt the AGN models of \citet{dale2014}  and introduce the
parameter `AGN fraction' in fitting. It turns out that in each region,
the AGN fraction to the SED is less than 10\%. Therefore, the
existence of AGN does not affect our results. 

\subsection{Influence of the extinction law}
\label{subsec:extcurve}

In Section \ref{subsec:physinterp}, we interpreted the different values
of the $E(B-V)_s^{\rm spec}$, $E(B-V)_s^{\rm SED}$, and $E(B-V)_g$ as the
results of a geometrical effect that the young and old stellar
populations may suffer from different dust attenuation because of
their different spatial distribution. In reality, the situation may be
more complicated. 
 
The derived color excess depends on the extinction law. For
simplicity and considering the starburst nature of Mrk848, we adopted
the dust extinction curve of \cite{calz2000} in the above
study. Here we test the influence of different extinction curves using
the CCM extinction curve of
\citet{cardelli1989}, and varying the $R_V'$ in Calzetti Law from 3.1
to 5.0. We find that the
conclusion is not affected by the different implementation of
extinction. The variation of the extinction curve is
very complicated and beyond the scope of this work. Here the
Calzetti extinction curve is a good approximation considering the
starburst nature of Mrk848.

\section{Star formation in Mrk848}
\label{sec:sf}

\subsection{Star formation history}
Measuring the SFH of a galaxy from its observed SEDs is not
trivial in practice. The best approach to providing 
unbiased results might be to model the SEDs with non-parametric SFHs
\citep{conroy2013}. However, the non-parametric SFH-reconstruction method requires not only data of very high quality,
 but also a good understanding of the dust attenuation.

\begin{figure*} 
\centering
\includegraphics[width=0.9\linewidth]{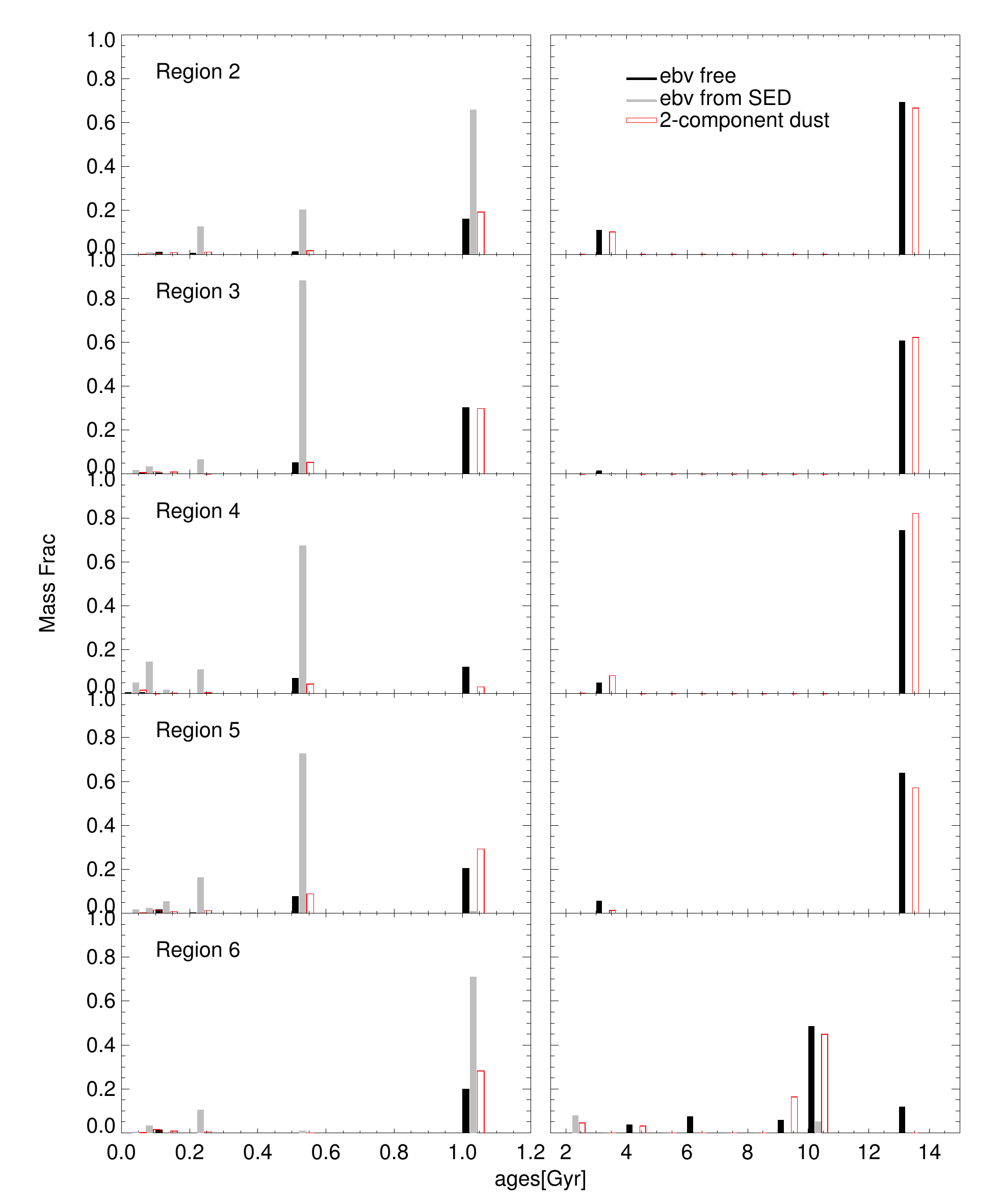}\\
\caption{Stellar populations of each region derived from pPXF,
  normalized to the total mass of each region, as a function of
  stellar ages. The black histograms show the results from pPXF fitting with
$E(B-V)_s$ as a free parameter. The gray 
histograms are derived from pPXF fitting of the spectra
using the $E(B-V)_s$ from the SED fitting. 
The red open histograms show the results given by using the
two-component dust model.} 
\label{fig:SFH}
\end{figure*}

In this work, the pPXF code was used to reconstruct the non-parametric
SFHs. We used two different
methods to treat the dust. First, we let the attenuation be a free 
parameter and let the pPXF fit the spectra, as described in
Section~\ref{subsec:specfitting}. Second, we corrected the
dust attenuation of the spectra 
using the $E(B-V)_s$ values obtained from the SED fitting in
  Section~\ref{subsec:sedfitting} and then
conducted the spectral fitting. The resulting 
SFHs are shown in Figure~\ref{fig:SFH}.  

The pPXF fitting with $E(B-V)_s$ as a free parameter gives a
large fraction of older stellar
populations and a very low fraction of 
stellar populations younger than 500~Myr. In contrast, the fitting of
the corrected spectra results in a very 
high fraction of young stellar populations and an extremely low fraction
of stellar populations older than 1~Gyr.
Neither of the two methods seem reliable considering that the
observed high fluxes both in UV/FIR and in NIR imply that these
regions should have considerable fractions of both young and old stellar
populations. Therefore, the former method
underestimates the fraction of the young stellar 
populations, whereas the latter overcorrects
the dust attenuation and
underestimates the fraction of the old 
stellar populations. 

According to the two-component dust model, the attenuation of the old
stellar populations and the young 
stellar populations should be treated differently. The young stellar
populations are more attenuated by 
dust than the old stellar populations. Therefore, because the above
two methods considered only one dust component which gives the
effective dust attenuation, over the whole region they overcorrect the dust
attenuation of the old stellar populations, and undercorrect that 
of young stellar populations. 

We then attempt to conduct the spectra fitting by adopting
the two-component dust model. We assume 
that the stellar populations younger than $t_{0}$ years old are more
attenuated than the older stellar populations, and that $E(B-V)_{young}$ is two times as great as $E(B-V)_{old}$. Letting $t_{0}$
change from 10~Myr to 50~Myr and $E(B-V)_{old}$ 
from 0.05 to 0.3, we search for the solution with the minimal
$\chi^2$. The resulting SFHs (open histograms in
Figure~\ref{fig:SFH})) show both the burst populations and 
the old stellar populations, consistent with the picture implied by
the high fluxes in both UV/IR and NIR. This is more reasonable than the
results given by the one-component dust model, where only a young or
old population exists.  

The SFHs can also be obtained from the SED fitting by CIGALE. Since we
adopt the assumption of an SFH with two exponentially decreasing
components for each region, the age
of the burst $t_{\rm burst}$, the duration of the burst $\tau_{\rm
  burst}$ and the mass fraction of the burst $f_{\rm burst}$ are used
to constrain the young stellar population, and the old stellar
population is constrained by the parameter $\tau_{\rm main}$,
indicating the time scale of the star formation (see
Sect.~\ref{subsec:sedfitting}). Table~\ref{tab:sedres} presents the SFH
parameters derived from the SED fitting.  

The UV-to-IR SED fitting provides good constraints on the dust attenuation of each region. However, since the SED fitting
assumes an analytical SFH, the real SFH, which may be quite
complicated and highly stochastic, cannot be well
reconstructed. Some studies have shown that it is difficult to
disentangle the different scenarios of the SFH because
of the degeneracy of the fits \citep[e.g.][and references
therein]{pforr2012,conroy2013,buat2014,buat2015,fernandez2015}. A
  recent work by
\citet{smith2015} showed that SFHs of isolated disk galaxies can be
recovered well using the Bayesian approach. However, they failed to
recover the SFH of mergers because the star formation in interacting
galaxies is much more complicated than in isolated disks. Our mock
analysis in Section~\ref{subsec:sedfitting}
(Figure~\ref{fig:comp_mock}) shows that 
the stellar mass and the SFR are well constrained by the SED
fitting. However, the parameters of the SFH
such as $f_{burst}$ and $t_{burst}$ are weakly constrained. 

Comparing the SFHs derived from the spectral fitting and the SED
fitting is not straightforward, because the spectral fitting gives
non-parametric and discrete SFHs while the SED fitting gives parametric
and continuous SFHs. Qualitatively, we find that the
star formation histories derived from the SED fitting are
consistent with the spectral fitting using the
two-component dust model. Both methods indicate that a significant
fraction ($\sim$15\%) of masses are contributed by stellar populations
less than $500$~Myr. This recent star formation may be triggered by
the galaxy interaction. Both methods
show that there is very recent star formation ($\sim$100~Myr) in one
of the core regions (R4). Instead the tail regions do not show a
significant fraction of stellar populations younger than $100$~Myr, but
contain more stellar populations with ages of $500$~Myr to $1$~Gyr. The
results may also indicate that the tail regions experience merger
induced starburst earlier than the core regions, which is consistent
with the scenario mentioned in Section~\ref{subsec:physinterp} that the gas
inflows from the tails to the center and cools down along the path.  

\begin{table}
  \centering
  \caption{Results of star formation history from broadband
    SED fitting.}
  \begin{adjustbox}{max width=\linewidth}
  \begin{tabular}{c c c c c c c}
    \hline\hline
            \multicolumn{7}{c}{Two exponentially-decreasing SFH} \\
            \hline
    Region  & $f_{burst}$ & $t_{burst}$ & $\tau_{burst}$ &
            $\tau_{main}$ & SFR  & log$M_{*}$\\
            &            & [Myr]      &   [Myr]   &
            [Myr]   &  [$M_{\odot}$yr$^{-1}$] & [$M_{\odot}$]\\
            \hline
    R1 & 0.06 & 408.5 & 1032.5 & 9858.2  &1.08  &  9.78\\ 
    R2 & 0.30 & 751.1 & 869.1 & 9512.5   &1.14  &  9.70\\
    R3 & 0.34 & 544.8 & 1118.2 & 9082.9  &11.35 &  10.28\\
    R4 & 0.26 & 96.0  & 975.9 & 9030.3  &76.45  &  10.45\\
    R5 & 0.37 & 707.5 & 965.9 & 9195.9  &1.41  &   9.53\\
    R6 & 0.20 & 617.0 & 965.9 & 9445.4  &1.20 &   9.63\\
    \hline
    \end{tabular}
    \end{adjustbox}
\label{tab:sedres}
\end{table}

\subsection{Star formation rate}
\begin{figure}
\centering
\includegraphics[width=9cm]{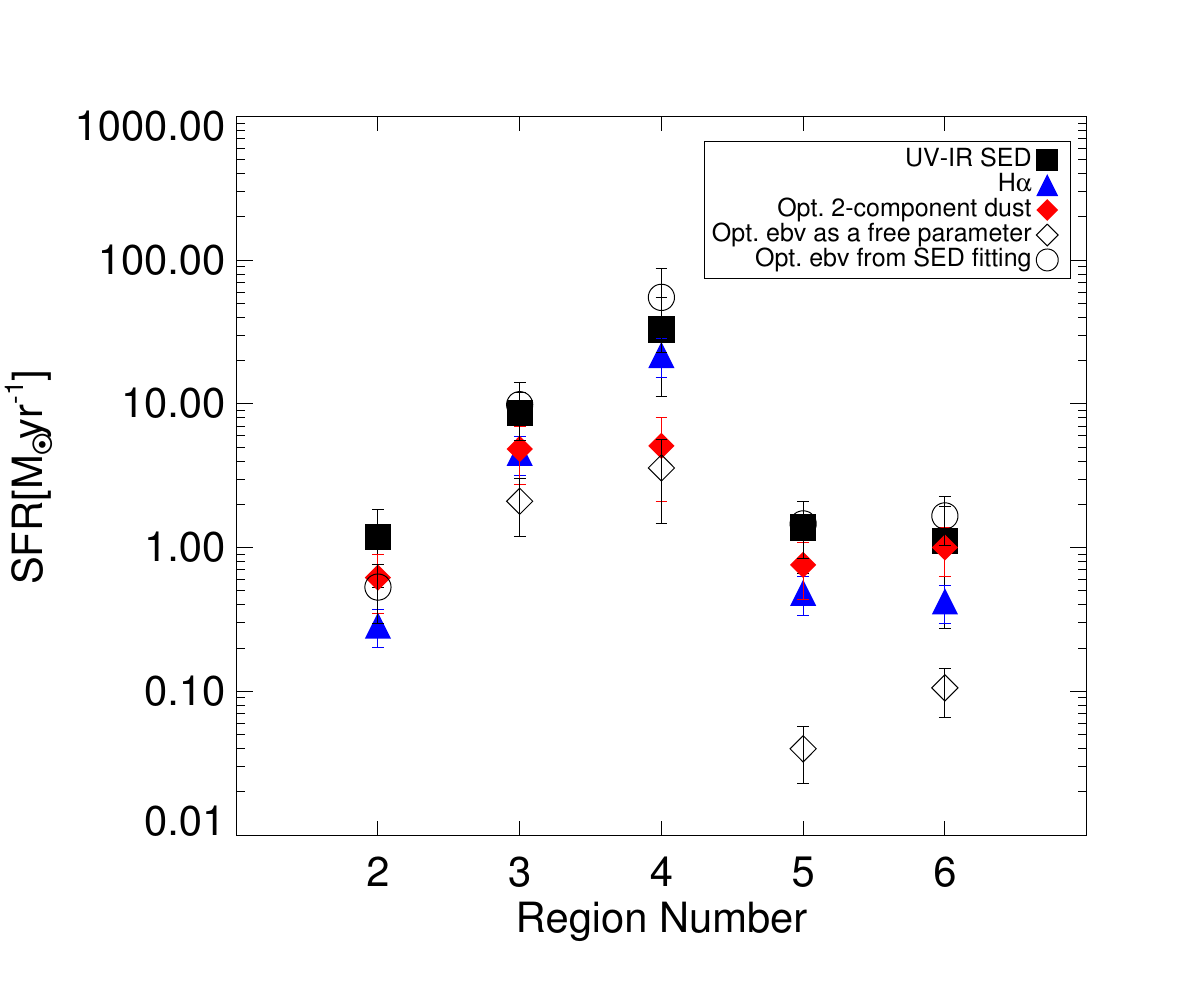}
\caption{Comparison of the SFRs derived from the UV-to-IR SED fitting
  (filled squares), the dust corrected ${\rm H}\alpha$ fluxes
  (filled triangles), the
  optical continuum stellar population fitting using the two-component
  dust model (filled diamonds), the
  optical continuum stellar population fitting with $E(B-V)$ as a free
  parameter (open diamonds), and the
  optical continuum stellar population fitting with $E(B-V)$ from
  UV-to-IR data (open circles).}
\label{fig:compare_sfr}
\end{figure}

Next, we compare the SFRs obtained with the two methods. Here
we take the SFR averaged in the recent 100~Myr.
Figure~\ref{fig:compare_sfr} shows the
comparison of the SFRs derived from the UV-to-IR SED fitting, the
optical continuum stellar population fitting and the ${\rm H}\alpha$
fluxes. If all comparing to the SFR estimated from the UV-to-IR
  SED method, the results from the optical continuum fitting (open
  diamonds) with
  $E(B-V)$ as a free parameter underestimate the SFR by $\sim$
  1~dex. The consistency is improved when the two-component
  dust model is applied (filled diamonds). However, the SFRs are still
  underestimated, especially for the R4 region, which has a very high
  SFR, implying that the SFRs in active star-forming regions are not
  well constrained by the optical continuum data only. A better
  consistency is obtained when the $E(B-V)$ information from UV-to-IR
  is introduced in the optical continuum fitting (open
  circles). We also calculate the SFRs using the dust corrected
  ${\rm H}\alpha$
  emission. We corrected the ${\rm H}\alpha$ luminosity using both the
  Balmer decrement and the $8$\micron~data with the relation
  provided in \citet{kennicutt2009} and
  \citet{ke2012}. Both corrections show similar results. 
Figure~\ref{fig:compare_sfr} shows that the dust corrected
${\rm~H}\alpha$ derived
SFRs (corrected using the $8$\micron~data) are lower than the
UV-to-IR SED derived SFRs. If we consider that
part of the H$\alpha$ could be contributed by the shock excitation
\citep{rich2011,rich2015}, the SFRs derived from H$\alpha$ should be even
lower. The discrepancy is partly due to that the IFU data do
not fully cover every region. Some pixels are missed because of the
limited size of the IFU and/or the bad data quality. We scale the average
value of the observed pixels to the whole region when calculating the
SFRs, which may account for part of the differences from the UV-to-IR
SED results
that are derived from the data with full coverage of Mrk848. 
Also, the wavelength coverage in our SED fitting is limited to
$8$\micron, which may produce large scatters in the estimated SFRs
\citep{calzetti2007}. Apart
 from these uncertainties, the lower SFR derived
 from the dust corrected H$\alpha$ is consistent with the
 two-component dust model with a large covering factor of the birth
 clouds \citep{wild2011}.

\section{Summary}
\label{sec:summary}

The plentiful data compiled both from the UV-to-IR multiwavelength
photometry and from the MaNGA IFU spectroscopy enable us to study the
galaxy merger Mrk848 in terms of the spatially resolved star formation properties. We divide the merger into six regions, among
which two regions belong to the core area where the star formation is
very intense and others belong to the tail area. In each region, we
use the UV-to-IR SED and the IFU spectrum respectively to analyze the dust
attenuation and the star formation of the merger.

The dust attenuation of stars derived from the UV-to-IR SED
$E(B-V)_s^{\rm SED}$ is consistent with that estimated from the UV/IR
indicators, including the infrared excess (IRX) and the UV slope
($\beta$). Furthermore, we test the reliability of the 
$E(B-V)_s^{\rm SED}$ with a set of mock SEDs generated by deviating
the best-fit
model SEDs, showing that the dust attenuation of the continuum can be
well constrained by the UV-to-IR broadband SED fitting.

However, the dust attenuation of stars derived from the optical
spectral fitting $E(B-V)_s^{\rm spec}$ is smaller than that from the
multi-band SED
fitting. When compared to the $E(B-V)_g$ obtained from the Balmer
  decrement, the ratio of the $E(B-V)_s^{\rm spec}$
to the $E(B-V)_g$ is 0.5, consistent with the local value, while the
$E(B-V)_s^{\rm SED}/E(B-V)_g$ is $\sim 1$. The difference between the results
from the UV-to-IR data and the optical data can be explained by the
model with two-component dust, in which the younger stellar
populations are embedded in the birth clouds and thus are attenuated
by both the dust from the surrounding birth clouds and the dust from
the ISM, whereas the older stellar
populations are only attenuated by the diffuse dust.
 
When adopting the two-component dust model in the spectral fitting,
the resulting SFH become qualitatively consistent with that from the
multi-band SED fitting. It might be a caution that the simple
one-component dust model
is not applicable for the mergers which are more complicated in the
distribution and the properties of the dust, and that the
two-component dust model is necessary to reconstruct the SFH of the
galaxy merger.

By investigating the SFH of the core and tail regions in Mrk848, we find
that both the spectral fitting and the SED fitting methods indicate
very recent starburst ($\sim100$~Myr) in core regions, especially in
region 4 which has the most intense star formation. The 
tail regions also show certain sign of recent starburst, but the burst
seems to have happened earlier ($\sim500$~Myr) than in the core
regions. The high SFRs in the core regions are consistent with the
interaction induced starburst.

Our results imply that,  it is a non-trivial task to break the
degeneracy between the stellar population and dust attenuation for the
merging galaxy. Information from multiwavelength observations from the
UV to the IR is necessary to constrain the dust attenuation and to
reconstruct the SFH.

\begin{acknowledgements}

We thank the anonymous referee for a thorough and constructive
  review. This work is supported by the
National Natural Science Foundation of
China (NSFC) with the Project Number of 11303070 (PI: FTY), 11433003
(PI:Chenggang Shu) and 11573050 (PI:SS). FTY is sponsored by Natural
Science Foundation of Shanghai (Project Number:
17ZR1435900) and the China Scholarship Council
  (CSC). C.J. acknowledges
support from the Natural Science Foundation of Shanghai
(No. 15ZR1446600), the National Natural Science Foundation of China
(NSFC, No. 11773051), and the CAS Key Research Program of Frontier
Sciences (No. QYZDB-SSW-SYS033). MAF is grateful
for financial support from CONICYT FONDECYT project No. 3160304. MB
was supported by MINEDUC-UA projects, code ANT 1655 and ANT 1656, and
FONDECYT project 1170618.

This research made use of Marvin, a core Python package and web
  framework for MaNGA data, developed by Brian Cherinka, José
  Sánchez-Gallego, and Brett Andrews. (MaNGA Collaboration, 2017).  
This work is based in part on observations made with the Spitzer
Space Telescope, which is operated by the Jet Propulsion Laboratory,
California Institute of Technology under a contract with NASA. 

Funding for the Sloan Digital Sky Survey IV has been provided by the
Alfred P. Sloan Foundation, the U.S. Department of Energy Office of
Science, and the Participating Institutions. SDSS acknowledges
support and resources from the Center for High-Performance Computing at
the University of Utah. The SDSS web site is www.sdss.org.

SDSS is managed by the Astrophysical Research Consortium for the
Participating Institutions of the SDSS Collaboration including the
Brazilian Participation Group, the Carnegie Institution for Science,
Carnegie Mellon University, the Chilean Participation Group, the
French Participation Group, Harvard-Smithsonian Center for
Astrophysics, Instituto de Astrofísica de Canarias, The Johns Hopkins
University, Kavli Institute for the Physics and Mathematics of the
Universe (IPMU) / University of Tokyo, Lawrence Berkeley National
Laboratory, Leibniz Institut für Astrophysik Potsdam (AIP),
Max-Planck-Institut für Astronomie (MPIA Heidelberg),
Max-Planck-Institut für Astrophysik (MPA Garching),
Max-Planck-Institut für Extraterrestrische Physik (MPE), National
Astronomical Observatories of China, New Mexico State University, New
York University, University of Notre Dame, Observatório Nacional /
MCTI, The Ohio State University, Pennsylvania State University,
Shanghai Astronomical Observatory, United Kingdom Participation Group,
Universidad Nacional Autónoma de México, University of Arizona,
University of Colorado Boulder, University of Oxford, University of
Portsmouth, University of Utah, University of Virginia, University of
Washington, University of Wisconsin, Vanderbilt University, and Yale
University. 
\end{acknowledgements}

\bibliographystyle{aa}
\bibliography{ref}

\begin{appendix}
\onecolumn
\section{Spectral fitting with pPXF for regions
  in Mrk848}

\begin{figure*}
\centering
\includegraphics[width=15cm]{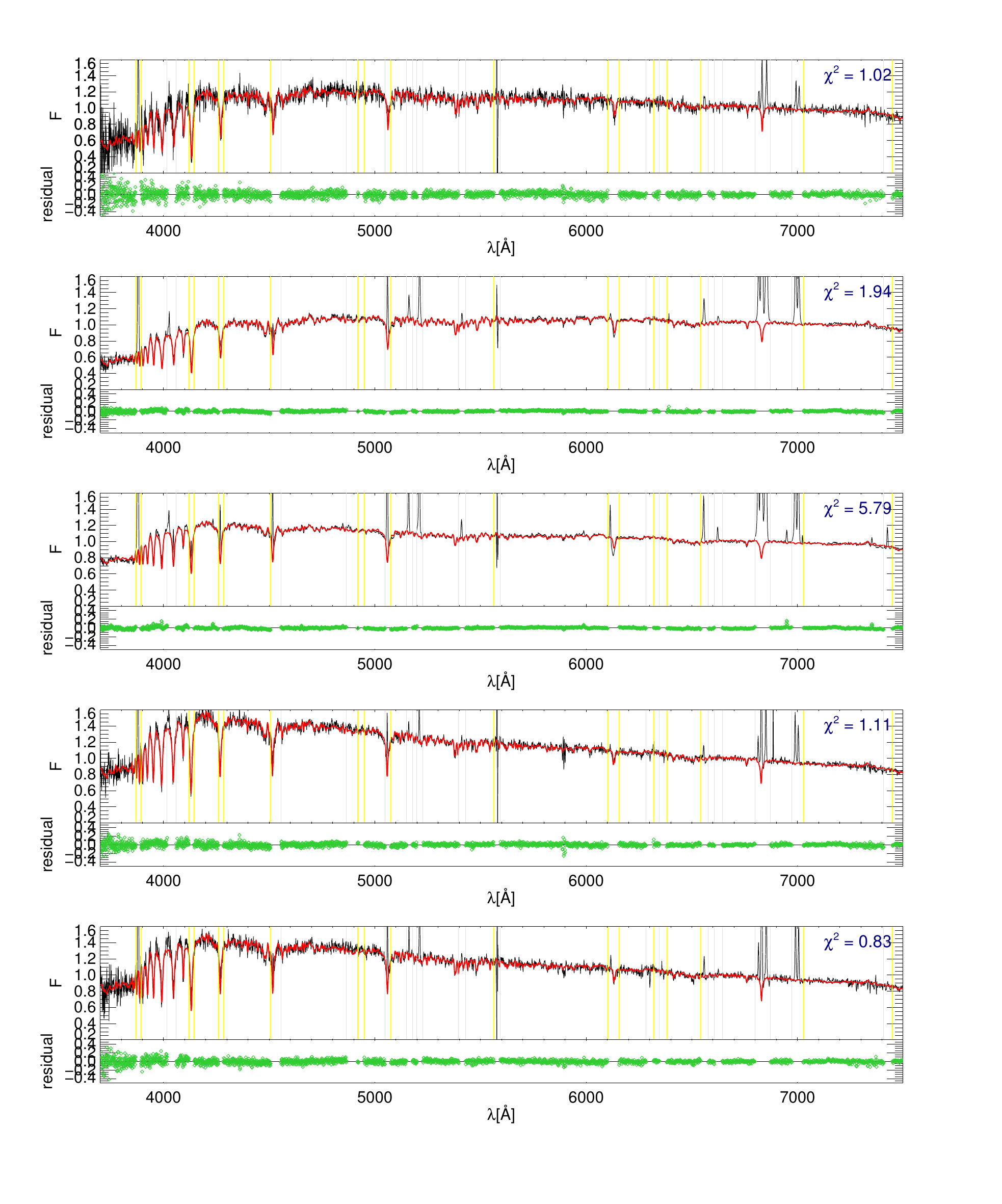}
\caption{Spectra (black lines), best-fit models, (red lines) and
  residuals (green dots) of pPXF fitting for
  the different regions in Mrk848 with
  the $E(B-V)$ as a free parameter. The yellow vertical
  lines show the masked emission line regions. }
\label{fig:spectra}
\end{figure*}

\begin{figure*}
\centering
\includegraphics[width=15cm]{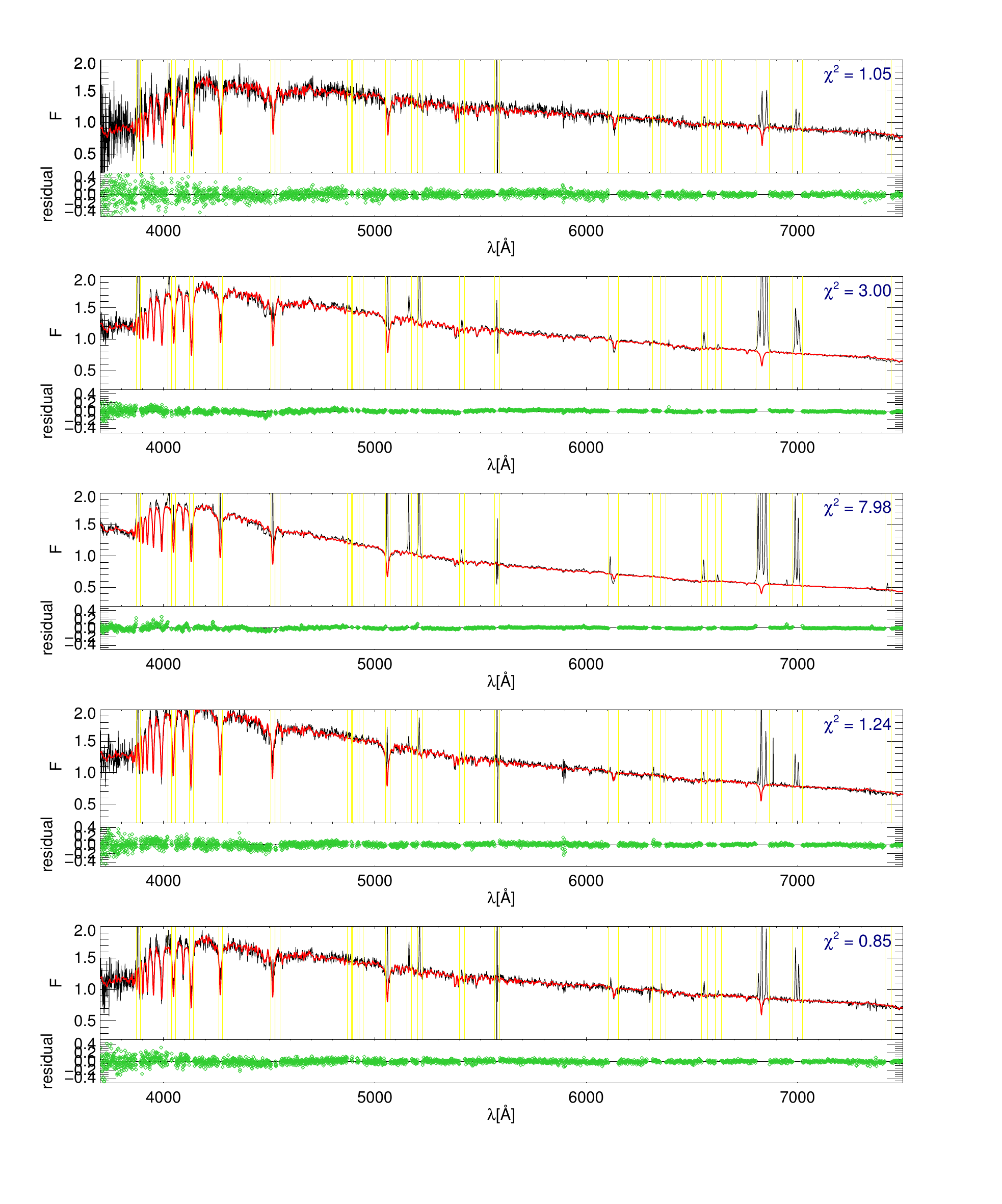}
\caption{Spectra, best-fit models, and residuals of pPXF fitting for
  the different regions in Mrk848 with
  the $E(B-V)$ derived from the UV-to-IR data. Symbols are the same as
in Figure \ref{fig:spectra}.}
\label{fig:spectra_att}
\end{figure*}

\begin{figure*}
\centering
\includegraphics[width=15cm]{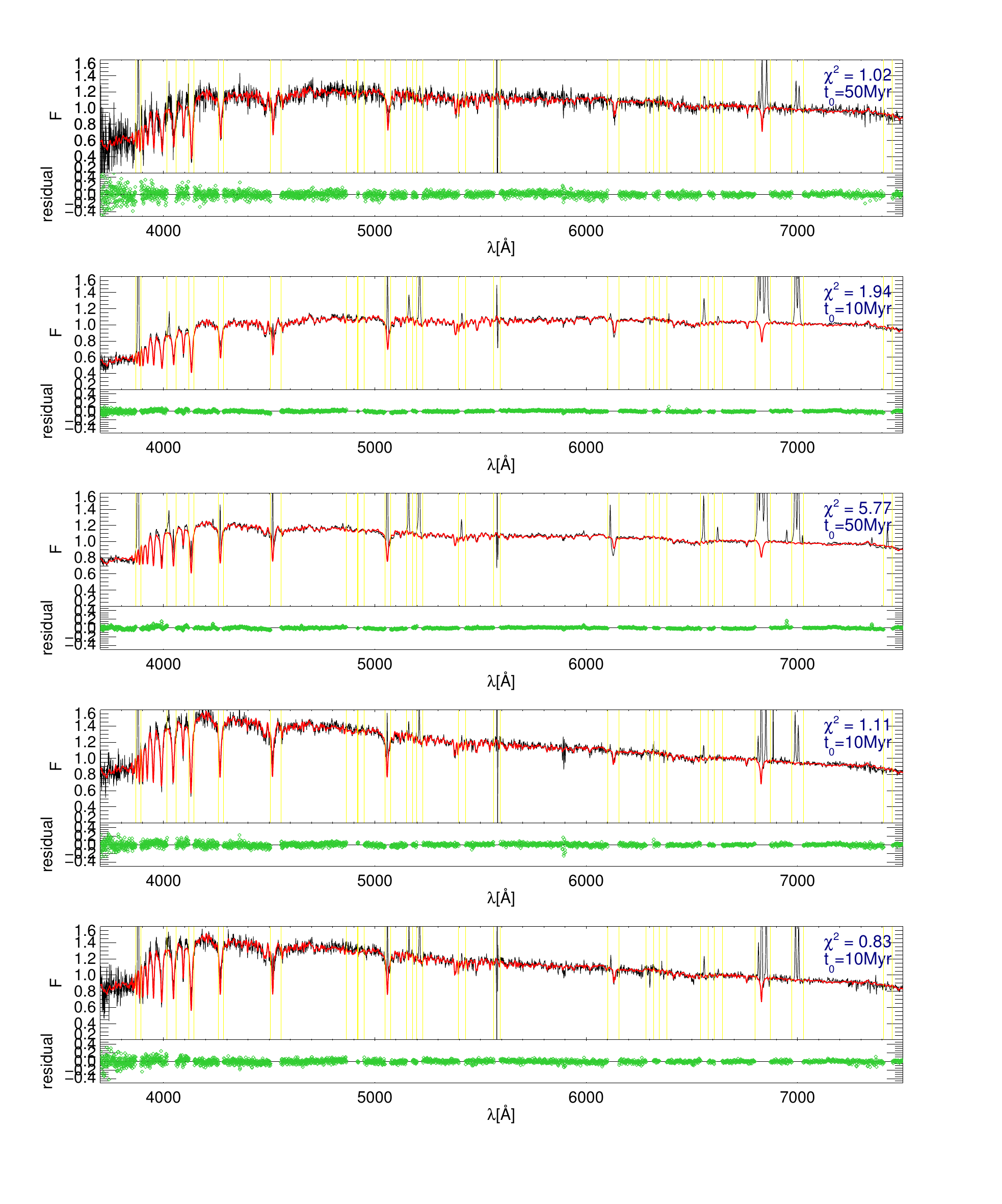}
\caption{Spectra, best-fit models, and residuals of pPXF fitting for
  the different regions in Mrk848 with the two-component dust model,
  assuming that the stellar populations less than $t_0$~Myr are
  attenuated twice of the stellar populations older than
  $t_0$~Myr. Symbols are the same as in Figure \ref{fig:spectra}.}
\label{fig:spectra_2pop}
\end{figure*}

We present the best-fit models and the residuals of the optical
spectral fitting with pPXF. Figure~\ref{fig:spectra} shows the results
assuming one-component dust and fitting the $E(B-V)$ as a free
parameter. Figure~\ref{fig:spectra_att} shows the results
assuming one-component dust and adopting the $E(B-V)$ from
UV-to-IR SED
fitting. Figure~\ref{fig:spectra_2pop} shows the fitting results
assuming the two-component dust model.

\end{appendix}

\end{document}